\pdfoutput=1
\documentclass{article}% Math and Physical Sciences Reference Style
\usepackage[scale=.715]{geometry}
\usepackage{authblk}
\usepackage{amsmath, amssymb, bm, breqn, cases, mathtools, wasysym}
\usepackage[bbgreekl]{mathbbol}
\usepackage{circuitikz, pgfplots, tikz}
\tikzset{>=latex}
\pgfplotsset{compat=1.17}
\usetikzlibrary{positioning}
\usepackage[font=small]{caption}
\usepackage[babel]{csquotes} %????
\usepackage[colorlinks,linkcolor=blue]{hyperref}

\newcommand*{\Sec}{sec.~}
\newcommand*{\Eq}{eq.~}
\newcommand*{\Eqs}{eqs.~}
\newcommand*{\Th}{Theorem~}

\newcommand*{\spec}{black}
\newcommand*{\latt}{black}
\newcommand*{\spac}{black}

\newcommand*{\functiondef}[5]{\begin{aligned} #1 : &&\hspace{-5pt} #2 &\,\to &&\hspace{-5pt} #3 \\ &&\hspace{-10pt} #4 &\,\mapsto &&\hspace{-5pt} #5 \end{aligned}}
\newcommand*{\di}{\mathrm{d}}
\newcommand*{\nnabla}{\textcolor{\spac}{\nabla}}
\DeclareMathOperator{\C}{C^*\!}
\DeclareMathOperator{\Div}{div}
\DeclareMathOperator*{\argmin}{argmin}
\makeatletter
\newcommand*\cdot@[2]{\mathbin{\vcenter{\hbox{\scalebox{#2}{$\m@th#1\bullet$}}}}}
\newcommand*\bcdot{\color{\latt}\mathpalette\cdot@{.55}\color{black}}
\makeatother

\DeclarePairedDelimiter\abs{\lvert}{\rvert}
\DeclarePairedDelimiter\pair{\langle}{\rangle}
\makeatletter \def\big{\bBigg@{1.25}} \makeatother

\newcommand*{\bszero}{\textcolor{\spec}{\boldsymbol{0}}}
\newcommand*{\bbzero}{\textcolor{\latt}{\mathbb{0}}}

\newcommand*{\bsone}{\textcolor{\spec}{\boldsymbol{1}}}
\newcommand*{\bbone}{\textcolor{\latt}{\mathbb{1}}}
\newcommand*{\bsa}{\textcolor{\spec}{\boldsymbol{a}}}
\newcommand*{\bba}{\textcolor{\latt}{\mathbb{a}}}
\newcommand*{\bb}{\mathrm{b}}
\newcommand*{\bbb}{\mathbb{b}}
\newcommand*{\bsc}{\textcolor{\spec}{\boldsymbol{c}}}
\newcommand*{\bbc}{\textcolor{\latt}{\mathbb{c}}}
\newcommand*{\bse}{\textcolor{\spec}{\boldsymbol{e}}}
\newcommand*{\bbe}{\textcolor{\latt}{\mathbb{e}}}
\newcommand*{\bsg}{\textcolor{\spec}{\boldsymbol{g}}}
\newcommand*{\bbg}{\textcolor{\latt}{\mathbb{g}}}
\newcommand*{\eell}{\textcolor{\spac}{\ell}}

\newcommand*{\ii}{\textcolor{\spac}{i}}
\newcommand*{\jj}{\textcolor{\spac}{j}}
\newcommand*{\bsm}{\textcolor{\spec}{\boldsymbol{m}}}
\newcommand*{\bbm}{\textcolor{\latt}{\mathbb{m}}}
\newcommand*{\bsn}{\textcolor{\spec}{\boldsymbol{n}}}
\newcommand*{\bbn}{\textcolor{\latt}{\mathbb{n}}}
\newcommand*{\bss}{\textcolor{\spec}{\boldsymbol{s}}}
\newcommand*{\bbs}{\textcolor{\latt}{\mathbb{s}}}
\newcommand*{\bbu}{\textcolor{black}{\mathbb{u}}}
\newcommand*{\bbv}{\textcolor{black}{\mathbb{v}}}
\newcommand*{\xx}{\textcolor{\spac}{x}}
\newcommand*{\bsy}{\textcolor{\spec}{\boldsymbol{y}}}
\newcommand*{\bsz}{\textcolor{\spec}{\boldsymbol{z}}}
\newcommand*{\bsalpha}{\textcolor{\spec}{\boldsymbol{\alpha}}}
\renewcommand*{\bbalpha}{\textcolor{\latt}{\mathbb{\alpha}}}
\newcommand*{\bsbeta}{\textcolor{\spec}{\boldsymbol{\beta}}}
\renewcommand*{\bbbeta}{\textcolor{\latt}{\mathbb{\beta}}}
\newcommand*{\bsgamma}{\textcolor{\spec}{\boldsymbol{\gamma}}}
\renewcommand*{\bbgamma}{\textcolor{\latt}{\mathbb{\gamma}}}
\newcommand*{\bsmu}{\textcolor{\spec}{\boldsymbol{\mu}}}
\renewcommand*{\bbmu}{\textcolor{\latt}{\mathbb{\mu}}}
\newcommand*{\bbB}{\mathbb{B}}
\newcommand*{\B}{\mathrm{B}}
\newcommand*{\bsD}{\textcolor{\spec}{\boldsymbol{D}}}
\newcommand*{\N}{\mathbb{N}}
\newcommand*{\cC}{\mathcal{C}}
\newcommand*{\cI}{\mathcal{I}}
\newcommand*{\cH}{\mathcal{H}}
\newcommand*{\cL}{\mathcal{L}}
\newcommand*{\cN}{\mathcal{N}}
\newcommand*{\cP}{\mathcal{P}}
\newcommand*{\Q}{\mathbb{Q}}
\newcommand*{\R}{\mathbb{R}}
\newcommand*{\T}{\mathbb{T}}
\newcommand*{\cT}{\mathcal{T}}
\newcommand*{\V}{\mathcal{V}}
\newcommand*{\Z}{\mathbb{Z}}

\title{A route to the hydrodynamic limit of a reaction-diffusion master equation using gradient structures}
\author[1]{Alberto Montefusco}
\author[1,2]{Christof Sch\"utte}
\author[1]{Stefanie Winkelmann}
\affil[1]{Mathematics of Complex Systems, Zuse-Institut Berlin,\protect\\Takustra\ss e 7, 14195 Berlin, Germany}
\affil[2]{Mathematics Institute, Freie Universit\"at Berlin,\protect\\Arnimallee 6, 14195 Berlin, Germany}

\begin{document}

\maketitle

\begin{abstract}
	The reaction-diffusion master equation (RDME) is a lattice-based stochastic model for spatially resolved cellular processes. It is often interpreted as an approximation to spatially continuous reaction-diffusion models, which, in the limit of an infinitely large population, may be described by means of reaction-diffusion partial differential equations (RDPDEs). Analyzing and understanding the relation between different mathematical models for reaction-diffusion dynamics is a research topic of steady interest. In this work, we explore a route to the hydrodynamic limit of the RDME which uses gradient structures. Specifically, we elaborate on a method introduced in [J. Maas, A. Mielke.: Modeling of chemical reactions systems with detailed balance using gradient structures. J.~Stat.~Phys.~(181), 2257--2303 (2020)] in the context of well-mixed reaction networks by showing that, once it is complemented with an appropriate limit procedure, it can be applied to spatially extended systems with diffusion. Under the assumption of detailed balance, we write down a gradient structure for the RDME and use the method to produce a gradient structure for its hydrodynamic limit, namely, for the corresponding RDPDE.
\end{abstract}

\section{Introduction}
Mathematical modeling and numerical simulation of reaction-diffusion processes is a research topic of steady interest. Application areas include many kinds of cellular processes like gene expression \cite{isaacson2006incorporating,winkelmann2016spatiotemporal}, neurotransmission \cite{ernst2021variance} % intracellular signaling
or enzyme kinetics \cite{straube2021stochastic}, %(biochemical reaction-diffusion systems), 
but also social dynamics of interacting agents such as innovation spreading or epidemics within a human population \cite{helfmann2021interacting,winkelmann2021mathematical}.
While, on the level of spatially well-mixed kinetics, the convergence of the stochastic jump process (characterized by the chemical master equation \cite{dtG92}) to the corresponding deterministic limit given by an ordinary differential equation is fully understood \cite{kurtz1970solutions}, an analogue analysis of the spatially resolved setting is not yet completed. 
Already the probabilistic formulation of particle-based reaction-diffusion dynamics -- as the root model to start with -- is a non-trivial issue because one has to couple the reaction kinetics to the continuous diffusion dynamics of a non-conserved and possibly unbounded number of particles. Here, applications of the Fock space formalism from quantum mechanics come into play, by which one can construct a characterizing evolution equation for particle-based reaction-diffusion dynamics using creation and annihilation operators  \cite{isaacson2020reaction,isaacson2020mean,del2021probabilistic}. 
Another approach, which will be used in this work,  is to discretize space by a regular Cartesian lattice and to replace the continuous spatial diffusion of particles by jumps between the lattice sites, which leads to the well-known reaction-diffusion master equation (RDME) \cite{hellander2015reaction,saI13,isaacson2009reaction}. 

Taking the hydrodynamic limit of the RDME means identifying a spatially continuous, deterministic, fluid-like model, described by a reaction-diffusion partial differential equation (RDPDE), which approximates the discrete model when the total number of particles is large. The hydrodynamic limit of reaction-diffusion systems has been studied in the last forty years \cite{DMFL85,DMFL86,pK88,dB94,aP00,mM18}, but only partial rigorous results, namely in simple specific cases, have been established so far. The main reason for the lack of results has to be traced in the nonlinearities that characterize the general reaction-diffusion systems. The first consequence of the presence of those nonlinearities is that a general theory for the well-posedness of RDPDEs is not yet available, except in the simplest cases, up to quadratic nonlinearities (cf.~\cite{mP10,jF15,LP20} for the most recent results). The second, related consequence is that proving that a hydrodynamic limit exists at all, i.e., \emph{compactness}, is challenging because the concentrations of the species may become locally unbounded \cite{mM18}.

From the non-rigorous, ``formal'' standpoint, the situation is rather clear instead. If the particle system is described by the RDME with mass-action-law stochastic reaction rates, the limit system, when the number of lattice sites goes to infinity and the order of magnitude of the particle numbers per lattice site is kept constant, is the RDPDE with mass-action-law deterministic reaction rates. The latter rates are averaged versions of the former, as we will see, with respect to a \emph{local-equilibrium} distribution.

Given compactness and local equilibrium, the hydrodynamic limit of interacting particle systems is usually proven at the level of the stochastic process \cite{KL99}. In this work, we take a different route %. Instead of showing the hydrodynamic limit at the level of the stochastic process, we use 
and, still not aiming at a full rigorous proof, apply a heuristic method by Maas and Mielke that uses \emph{gradient structures} \cite{MM20}. These are geometric structures that were introduced in both physical \cite{mG85,pjM86} and mathematical \cite{DGMT80,CV90} contexts to model dissipative phenomena, in particular in the literature of nonequilibrium thermodynamics \cite{lO31a,GO97}. To prove the hydrodynamic limit or other results, one works with the gradient structures associated with the system rather than directly with the evolution equations.

With the use of gradient structures, we aim to pave the way to (i) a rich thermodynamic insight in reaction-diffusion systems, (ii) a possible alternative scheme for rigorous proofs of their hydrodynamic limit, (iii) a general framework for the construction of hybrid models, namely models where part of the system is treated stochastically, partly deterministically. An essential requirement for the use of gradient structures is the presence of detailed balance \cite{aM11,MM20}.  

The method of Maas and Mielke, as introduced in \cite{MM20}, essentially transfers the gradient structures with the help of a local-equilibrium assumption. While this method works for well-mixed chemical reactions, for spatially extended systems this is not sufficient to get the desired result, but has to supplemented with a limit procedure.

The plan of the paper is as follows. In \Sec\ref{sec:RD}, we describe the main physical and mathematical elements to describe reaction-diffusion systems, and in particular the stochastic model (RDME) in \Sec\ref{sec:RDME} and the deterministic model (RDPDE) in \Sec\ref{sec:RDPDE}. In \Sec\ref{sec:GS}, we introduce the notion of a gradient structure and formulate how it is applied to both the RDME and the RDPDE. We then use the method of Maas and Mielke, in \Sec\ref{sec:limit}, to show that the RDPDE is indeed the hydrodynamic limit of the RDME. In the conclusions (\Sec\ref{sec:conclusion}), we summarize the results and suggest the main future steps.

\section{Reaction-diffusion systems}\label{sec:RD}
The systems we study in this work are characterized by the superposition of two different phenomena. When a substance diffuses in a container in the absence of external forces and cross effects with other dissipative phenomena, it tends to occupy the whole container by transferring from regions with higher concentrations to regions with lower concentrations. In this paper, we assume that the system consists of different substances that diffuse independently from each other. Furthermore, we suppose that even the single particles of the same species do not feel each other. The motion of each species is thus characterized by some diffusion coefficient or tensor. To avoid considering boundary conditions, the space where the motion takes place is a torus in $d$ dimensions.

A reaction models the sudden transformation of several species into different ones. This model is usually a coarse-grained picture of a complex system, where the effect of multiple interactions in a high-dimensional space gives rise to preferred, metastable states where the system spends most of its time. These preferred states are exactly what we refer to as ``species'', which are thus not necessarily chemical species, but may just be different important configurations of the same molecule, for instance. The set of all possible reactions is encoded by a \emph{reaction network} of the type
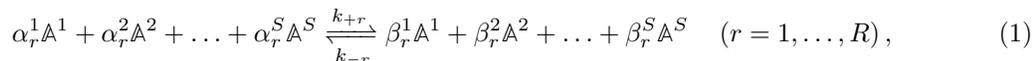
\begin{equation}\label{network}
	\alpha^1_r \mathbb{A}^1 + \alpha^2_r \mathbb{A}^2 + \ldots + \alpha^S_r \mathbb{A}^S \xrightleftharpoons[k_{-r}]{k_{+r}} \beta^1_r \mathbb{A}^1 + \beta^2_r \mathbb{A}^2 + \ldots + \beta^S_r \mathbb{A}^S \, \quad (r = 1, \ldots, R) \, ,
\end{equation}
constituted of $S$ species and $R$ reactions. The symbol~$\mathbb{A}^s$ indicates a species, $\bsalpha_r \coloneqq (\alpha^s_r)_{s=1,\ldots,S} \coloneqq (\alpha^1_r, \alpha^2_r, \ldots, \alpha^S_r)$ and $\bsbeta_r \coloneqq (\beta^s_r)_{s=1,\ldots,S}$, where $\alpha^s_r, \beta^s_r \in \N$ are (possibly zero) stoichiometric coefficients for the species~$s$ and the reaction~$r$, and $k_{+r}$ and $k_{-r}$ are the forward and backward reaction rates for the reaction~$r$. We will specify later the role of $k_{+r}$ and $k_{-r}$. When a reaction~$r$ occurs, the species~$s$ is changed by $\beta^s_r - \alpha^s_r$ unities. Furthermore, we introduce the notations, for any vectors $\bsm, \bsgamma \in \R^S$,
\begin{equation}\label{notation}
	\abs{\bsgamma} := \sum\limits_{s = 1}^S \gamma^s \, , \qquad \bsm^{\bsgamma} := \prod\limits_{s = 1}^S (m^s)^{\gamma^s} \, , \qquad \bsm! := \prod\limits_{s = 1}^S m^s! \, .
\end{equation}

By their combination, reaction and diffusion influence each other. On the one hand, in a certain region of the torus, reaction may decrease the amount of a certain species, which may however be transferred to that region by diffusion and thus be able to react again. We say that reactions are \emph{diffusion-limited} in this situation. On the other hand, diffusion of a species can occur only if that species is produced by reactions.

What may sound surprising is that both phenomena, in the stochastic model of this paper, can be thought of in exactly the same way: both diffusion and reaction constitute ``reaction'' phenomena in a proper abstract sense. What especially marks their difference does not lie in their inherent nature, but in the rate at which they occur. The difference in the scalings causes the deterministic limits of the stochastic systems to be of two different natures: one, indeed, of diffusive type, and the other one of reaction type.

\subsection{The reaction-diffusion master equation (RDME)}\label{sec:RDME}
By the expression \emph{reaction-diffusion master equation}, we mean a stochastic model where reaction and diffusion occur on a lattice according to a continuous-time Markov process. Diffusion is the part of the process by which particles jump from a lattice site to the neighboring ones. Reaction is a phenomenon that involves only one lattice site. In the following sections we present the two components of the stochastic process separately, and in the last section we see that, from an abstract viewpoint, they can be described essentially in the same way.

Before doing that, we introduce some further notation. The lattice is a $d$-dimensional space of $N^d$ points that are denoted by $\ii \coloneqq (i_\ell)_{\ell=1,\ldots,d}$ with $i_\ell \in [0, N]$; we silently assume periodic boundary conditions, namely that the point with $i_\ell = N$ is the same as the point with $i_\ell = 0$. The space is thus the discrete torus $\Z_N^d \coloneqq \Z^d / (N \Z)^d$. At each lattice point~$\ii \in \Z_N^d$, and for each species, we count the number of particles. The collection of all particle numbers at all points is denoted by $\bbn \coloneqq (\bsn_{\ii})_{\ii \in \Z_N^d} \in (\N^S)^{Z^d_N} \eqqcolon \cN$ with $\bsn_{\ii} \coloneqq (n^s_{\ii})_{s=1,\ldots,S} \in \N^S$ and $n^s_{\ii} \in \N$. We define
\begin{equation}
	\bbm! \coloneqq \prod\limits_{\ii \in \Z_N^d} \bsm_{\ii}! \, .
\end{equation}

Finally, the state of the RDME is described by the probabilities of each configuration $\bbn$, which we indicate by $u_{\bbn} \in \R_+$, and we collect all of these in the infinite-dimensional vector $\bbu$, an element of the space of probability measures on $\cN$, which we denote by $\cP(\cN)$ and may be identified with the space $\Bigl\{ \bbu \in [0, 1]^{\cN} \Bigm\vert \sum\limits_{\bbn \in \cN} \bbu_\bbn = 1 \Bigr\}$.

\subsubsection*{Reaction}
At each lattice site, there are $R$ possible reactions according to the network~\eqref{network}, and the particle numbers jump according to certain transition rates that can be modeled in various ways depending on the reaction types. In this paper, we assume that the rates are modeled according to the \emph{law of mass action}. The law-of-mass-action transition rates for a well-mixed system specify the \emph{chemical master equation} \cite{dtG92}.

Let us introduce the mass-action-law stochastic factors, for a generic $\bsgamma \in \N^S$,
\begin{equation}\label{factors}
	\functiondef{\B^\bsgamma}{\Z^S}{\Q\,,}{\bsn}{
		\begin{dcases*}
			\frac{\bsn!}{(\bsn - \bsgamma)!} & for $\bsn - \bsgamma \in \N^S$, \\
			0  & for $\bsn - \bsgamma \notin \N^S$.
	\end{dcases*}}
\end{equation}
This definition automatically assigns a zero rate to the reactions that do not have a sufficient number of particles to occur.

A forward reaction~$r$ makes the state of the system at position $\ii$ jump
\begin{subequations}\label{rateR}
\begin{equation}
	\text{from} \quad \bsn_\ii \quad \text{to} \quad \bsn_\ii - \bsalpha_r + \bsbeta_r \quad \text{with rate} \quad k_{+r} \, \B^{\bsalpha_r}\!(\bsn_{\ii})
\end{equation}
and a backward reaction
\begin{equation}
	\text{from} \quad \bsn_\ii \quad \text{to} \quad \bsn_\ii + \bsalpha_r - \bsbeta_r  \quad \text{with rate} \quad k_{-r} \, \B^{\bsbeta_r}\!(\bsn_{\ii}) \, ,
\end{equation}
\end{subequations}
with $k_{+r}, k_{-r} > 0$.

\subsubsection*{Diffusion}
Diffusion is modeled, at the stochastic level, as a jump process that transfers one particle from a lattice site to one of its neighbors. Let us introduce~$\{\eell\}_{\ell=1,\ldots,d}$, a basis of unit vectors of $\R^d$. A forward diffusion event in such a direction for the species $s$ moves a particle between two neighboring positions, namely it changes the state of the system at the positions $\ii$ and $\ii + \eell$
\begin{subequations}
\begin{equation}
	\text{from} \quad \begin{dcases} n^s_\ii \\ n^s_{\ii + \eell} \end{dcases} \quad \text{to} \quad \begin{dcases} n^s_\ii - 1 \\ n^s_{\ii + \eell} + 1 \end{dcases} \quad \text{with rate} \quad N^2 D^{s, +\eell}_\ii \, n^s_{\ii} \, ,
\end{equation}
and a backward event
\begin{equation}
	\text{from} \quad \begin{dcases} n^s_\ii \\ n^s_{\ii + \eell} \end{dcases} \quad \text{to} \quad \begin{dcases} n^s_\ii + 1 \\ n^s_{\ii + \eell} - 1 \end{dcases} \quad \text{with rate} \quad N^2 D^{s, -\eell}_{\ii+\eell} \, n^s_{\ii+\eell} \, ,
\end{equation}
\end{subequations}
with $D^{s, +\eell}_{\ii}, D^{s, -\eell}_{\ii} \geq 0$ at every $\ii$. Once we introduce
\begin{equation*}
	\bse^s \, , \, \text{the canonical unit vector in } \R^S \text{ in the direction } s \, ,
\end{equation*}
we may reformulate the transitions as
\begin{subequations}\label{rateD}
\begin{align}
	&\text{from} \quad \begin{dcases} \bsn_\ii \\ \bsn_{\ii + \eell} \end{dcases} \quad \text{to} \quad \begin{dcases} \bsn_\ii - \bse^s \\ \bsn_{\ii + \eell} + \bse^s \end{dcases} \quad \text{with rate} \quad N^2 D^{s, +\eell}_\ii \, \B^{\bse^s}\!(\bsn_\ii) \quad \text{and} \\
	&\text{from} \quad \begin{dcases} \bsn_\ii \\ \bsn_{\ii + \eell} \end{dcases} \quad \text{to} \quad \begin{dcases} \bsn_\ii + \bse^s \\ \bsn_{\ii + \eell} - \bse^s \end{dcases} \quad \text{with rate} \quad N^2 D^{s, -\eell}_{\ii+\eell} \, \B^{\bse^s}\!(\bsn_{\ii+\eell}) \, .
\end{align}
\end{subequations}
From the last formulation of the rates, we may already notice how reaction and diffusion have essentially the same structure in terms of the functions $\B$: diffusion is a special case for first-order reactions, which have linear rates. We recognize an even more elegant joint formulation in the next section.

\subsubsection*{Reaction-diffusion}
Let us define the functions on the lattice
\begin{equation}
	(\delta_{\ii})_{\jj} \coloneqq
	\begin{dcases*}
		1 & if $\jj = \ii$, \\
		0 & if $\jj \neq \ii$,
	\end{dcases*} \qquad
	\bbalpha^r_{\ii} \coloneqq \bsalpha_r \, \delta_{\ii} \, , \qquad \bbbeta^r_{\ii} \coloneqq \bsbeta_r \, \delta_{\ii} \, , \qquad \bbe^s_\ii \coloneqq \bse^s \, \delta_\ii \, ,
\end{equation}
and the rates
\begin{equation}
	\bbB^{\bbgamma}(\bbn) \coloneqq \prod\limits_{\ii \in \Z_N^d} \B^{\bsgamma_{\ii}}(\bsn_\ii) =
	\begin{dcases*} \prod\limits_{\ii \in \Z_N^d} \frac{\bsn_{\ii}!}{(\bsn_{\ii} - \bsgamma_{\ii})!} \eqqcolon \frac{\bbn!}{(\bbn - \bbgamma)!} & if $\bbn - \bbgamma \in \cN$, \\
		0 & otherwise. \end{dcases*}
\end{equation}
The transitions given in \eqref{rateR}-\eqref{rateD} may then be re-expressed as follows:
\begin{align*}
	&\text{from} \quad \bbn \quad \text{to} \quad \bbn - \bbalpha^r_{\ii} + \bbbeta^r_{\ii} &&\hspace{-60pt} \text{with rate} \quad k_{+r} \, \bbB^{\bbalpha^r_{\ii}}\!(\bbn) \, , \\
	&\text{from} \quad \bbn \quad \text{to} \quad \bbn + \bbalpha^r_{\ii} - \bbbeta^r_{\ii} &&\hspace{-60pt} \text{with rate} \quad k_{-r} \, \bbB^{\bbbeta^r_{\ii}}\!(\bbn) \, , \\
	&\text{from} \quad \bbn \quad \text{to} \quad \bbn - \bbe^s_{\ii} + \bbe^s_{\ii+\eell} &&\hspace{-60pt}\text{with rate} \quad N^2 D^{s, +\eell}_\ii \, \bbB^{\bbe^s_{\ii}}\!(\bbn) \, , \ \text{and} \\
	&\text{from} \quad \bbn \quad \text{to} \quad \bbn + \bbe^s_{\ii} - \bbe^s_{\ii+\eell} &&\hspace{-60pt} \text{with rate} \quad N^2 D^{s, -\eell}_{\ii+\eell} \, \bbB^{\bbe^s_{\ii+\eell}}(\bbn) \, .
\end{align*}
All of the rates are thus in the form $k \, \bbB^\bbgamma(\bbn)$, where $k$ is a rate constant and $\bbgamma$ is a generalized vector of stoichiometric coefficients: this is, instead of a vector in $\R^S$, a vector in $\cN = (\R^S)^{\Z^d_N}$. In the standard RDME that we consider in this paper, any of these generalized vectors have non-zero entries only for one lattice site $\ii$, meaning that reactants and products of a reaction are located at the same site. In some models, however, reactions are allowed between different lattice sites \cite{saI13}.

The reaction-diffusion stochastic system evolves according to the Kolmogorov forward equation associated with the transition rates that have just been defined, namely
\begin{subequations}\label{RDME}
	\begin{equation}
		\dot{\bbu}_t = \sum\limits_{\ii \in \Z_N^d} \sum\limits_{r=1}^R \mathfrak{R}^r_\ii(\bbu_t) + \sum\limits_{\ii \in \Z_N^d} \sum\limits_{s=1}^S \sum\limits_{\ell=1}^d \mathfrak{D}^{s, \eell}_\ii(\bbu_t) \, ,
	\end{equation}
	where the vector fields $\mathfrak{R}^r_\ii$ and $\mathfrak{D}^{s, \eell}_\ii$ have components
	\begin{align}
		\bigl(\mathfrak{R}^r_\ii(\bbu)\bigr)_{\bbn} &\coloneqq k_{+r} \bigl[ \B^{\bsalpha_r}\!(\bsn_{\ii} + \bsalpha_r - \bsbeta_r) \, u_{\bbn + \bbalpha^r_{\ii} - \bbbeta^r_{\ii}} - \B^{\bsalpha_r}\!(\bsn_{\ii}) \, u_{\bbn} \bigr] \nonumber \\
		&\hspace{3.7pt} + k_{-r} \bigl[ \B^{\bsbeta_r}\!(\bsn_{\ii} - \bsalpha_r + \bsbeta_r) \, u_{\bbn - \bbalpha^r_{\ii} + \bbbeta^r_{\ii}} - \B^{\bsbeta_r}\!(\bsn_{\ii}) \, u_{\bbn} \bigr] \, , \label{reactionVectorField} \\[2pt]
		\bigl(\mathfrak{D}^{s, \eell}_\ii(\bbu)\bigr)_\bbn &\coloneqq N^2 D^{s,+\eell}_\ii \bigl[ (n^s_\ii + 1) \, u_{\bbn + \bbe^s_\ii - \bbe^s_{\ii+\eell}} - n^s_\ii \, u_\bbn \bigr] \nonumber \\
		&\hspace{3.7pt} + N^2 D^{s,-\eell}_{\ii+\eell} \bigl[ (n^s_{\ii+\eell} + 1) \, u_{\bbn - \bbe^s_\ii + \bbe^s_{\ii+\eell}} - n^s_{\ii+\eell} \, u_\bbn \bigr] \, . \label{diffusionVectorField}
	\end{align}
\end{subequations}
Eq.~\eqref{RDME} is the RDME.

\subsection{The reaction-diffusion partial differential equation (RDPDE)}\label{sec:RDPDE}
When the size~$N$ of the system is sufficiently large and the total particle numbers scales accordingly, a deterministic model may constitute an accurate description of the dynamics of reaction-diffusion systems. The state of the system is described by $S$ fields of concentrations on the unit torus $\T^d = \R^d/\Z^d$, namely $[0, \infty)^S$-valued functions (or measures)~$\bbc$ on $\T^d$ (we write $\bbc \in \cC$) such that
\begin{equation*}
	\int_\V c^s(\xx) \, \di \xx = \#\{\text{particles of species } s \text{ in } \V\} \, ,
\end{equation*}
where $\V$ is a measurable subset of $\T^d$ and $\di \xx$ is the volume element on $\T^d$. Similarly to the discrete case, we also write $\bsc(\xx) = \bigl(c^s(\xx)\bigr)_{s=1,\ldots,S}$.

\subsubsection*{Reaction}
At every point in space, $R$ reactions may happen. By analogy with the stochastic factors $\B$ defined in \eqref{factors}, let us introduce the mass-action-law deterministic factors
\begin{equation}
	\bb^\bsgamma(\bsc) \coloneqq \bsc^\bsgamma \, .
\end{equation}
A forward reaction moves the state of the system at position $\xx$
\begin{equation}
	\text{in the direction}\quad \bsbeta_r - \bsalpha_r \quad\text{with rate} \quad k_{+r} \, \bb^{\bsalpha_r}\!\bigl(\bsc(\xx)\bigr) \, ,
\end{equation}
and a backward reaction
\begin{equation}
	\text{in the direction}\quad \bsalpha_r - \bsbeta_r \quad\text{with rate} \quad k_{-r} \, \bb^{\bsbeta_r}\!\bigl(\bsc(\xx)\bigr) \, .
\end{equation}

Furthermore, we define the factors
\begin{equation}
	\bbb^\bbgamma(\bbc) \coloneqq \bbc^\bbgamma \coloneqq \prod\limits_{\xx \in \T^d} \bsc(\xx)^{\bsgamma(\xx)} = \prod\limits_{\xx \in \T^d} \bb^{\bsgamma(\xx)}\!\bigl(\bsc(\xx)\bigr) \, ,
\end{equation}
and reformulate the rates as $k_{+r} \, \bbb^{\bbalpha^r_{N \xx}}(\bbc)$ and $k_{-r} \, \bbb^{\bbbeta^r_{N \xx}}(\bbc)$.

\subsubsection*{Diffusion}
The species are assumed to diffuse independently from each other with the well-known dynamics of the form  
\begin{equation}
	\Div\bigl( D^{s}(\xx) \, \nnabla c^s(\xx) \bigr) \, ,
\end{equation}
where $D^s(\xx)$ is symmetric and positive semidefinite, $\nnabla$ is the gradient operator on $\T^d$, and $\Div$ the divergence operator on $\T^d$. For a more condensed notation, we also define
\begin{equation}
	\bsD(\xx) \, \nnabla \bsc(\xx) \coloneqq \left( D^s(\xx) \, \nnabla c^s(\xx) \right)_{s = 1, \ldots, S}
\end{equation}
and, for $\bsm \in \R^{d \times S}$,
\begin{equation}\label{divergence}
	\Div\bsm \coloneqq \bigl( \Div m^s \bigr)_{s = 1, \ldots, S} \qquad \text{with } m^s \in \R^d \, .
\end{equation}

\subsubsection*{Reaction-diffusion}

The two components combine into the RDPDE
\begin{equation}
	\dot{\bsc}_t(\xx) = \sum\limits_{r=1}^R \left( k_{+r} \, \bsc_t(\xx)^{\bsalpha_r} - k_{-r} \, \bsc_t(\xx)^{\bsbeta_r} \right) \! (\bsbeta_r - \bsalpha_r) + \Div\bigl(\bsD(\xx) \, \nnabla \bsc_t(\xx) \bigr) \, ,
\end{equation}
also written as
\begin{subequations}\label{RDPDE}
	\begin{equation}
		\dot{\bbc}_t = \sum\limits_{r=1}^R \mathfrak{r}^r(\bbc_t) (\bsbeta_r - \bsalpha_r) - \Div \mathfrak{d}(\bbc_t)
	\end{equation}
	with
	\begin{equation}
		\mathfrak{r}^r(\bbc)(\xx) \coloneqq k_{+r} \, \bsc(\xx)^{\bsalpha_r} - k_{-r} \, \bsc(\xx)^{\bsbeta_r} \quad \text{and} \quad \mathfrak{d}^s(\bbc)(\xx) \coloneqq - \bsD(\xx) \, \nnabla \bsc(\xx) \, .
	\end{equation}
\end{subequations}

\section{Gradient structures for the RDPDE and the RDME}\label{sec:GS}
\subsection{Gradient structures in a nutshell}
\begin{figure}[t]
	\centering
	\begin{circuitikz}
		\draw node[eground, rotate=-90] {} (0, 0) to[spring, l=$E$, i_>=$z$] (1.5, 0) to[damper, l=\raisebox{3.25pt}{$\Psi$}] (3, 0) node[eground, rotate=90] {};
	\end{circuitikz}
	\caption{A typical potential energy is $E(z) = \frac{1}{2} k z^2$, where $k$ is the stiffness constant, and the simplest dissipation potential is $\Psi(z, \dot{z}) = \frac{1}{2} \mu \dot{z}^2$, where $\mu$ is the friction coefficient. These choices lead to the linear spring force $-k z$ and the linear damping $-\mu \dot{z}$. The force balance thus yields the linear ODE $\mu \dot{z}_t + k z_t = 0$.}
	\label{fig:spring-damper}
\end{figure}
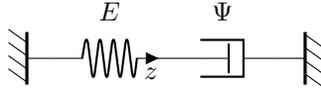

When detailed balance is satisfied (cf.~sec.~\ref{sec:DB}), both the RDME and the RDPDE can be given the form of a \emph{gradient flow}. This means that their dynamics may be expressed in terms of the balance of forces
\begin{equation}\label{GF1}
	- \di E(z_t) - \partial_{\dot{z}}\Psi(z_t, \dot{z}_t) = 0 \, ,
\end{equation}
where $\di$ and $\partial$ indicate the total and partial derivative operators. The first force is a potential restoring force and the second one is a frictional (or ``thermodynamic'') force. Any relation~$\xi = K(z, \dot{z})$ between the rate~$\dot{z}$ and the frictional force~$\xi$ is usually referred to as a \emph{kinetic relation}. A \emph{gradient structure} expresses such a kinetic relation as the derivative of a dissipation potential with respect to the rate~$\dot{z}$. The simplest example of the balance of forces is shown in \figurename~\ref{fig:spring-damper}.

Gradient structures constitute a model of dissipation for many physical systems and processes, such as (complex-) fluid dynamics, chemical reactions, heat conduction, rarefied gas dynamics \cite{hcO05,maP14,PKG18}. At the mathematical level, the theory of gradient structures is developing towards a broader and broader generality: the case of quadratic dissipation potentials is very well developed and allows us to treat evolution equations in abstract metric spaces \cite{AGS08}, and in particular in spaces of probability measures. The non-quadratic situation, to date, has not reached such level of abstraction (the latest attempt of a general theory in nonlinear situations is made in \cite{PRST20}). Since our case falls under this second category, we necessarily introduce the mathematical objects in a heuristic way.

We denote the state space by~$Z$, the space of \emph{rates} at the point~$z$ by~$T_zZ$ (the tangent space at $z$), and the space of \emph{forces} at the point~$z$ by $T^*_zZ$ (the cotangent space at $z$). The disjoint union of all tangent spaces is the tangent bundle~$TZ$, and the one of all cotangent spaces is the cotangent bundle~$T^*Z$.

In this paper, a \emph{gradient structure} on the (possibly infinitely dimensional) state space~$Z$, is a pair~$(E, \Psi)$ of
\begin{itemize}
	\item a \emph{driving function} $E : Z \to \R$, a continuously differentiable function;
	\item a \emph{dissipation potential} $\Psi : TZ \to [0, \infty]$ that, for every $z \in Z$, has the following properties
	\begin{enumerate}
		\item $\Psi(z, \cdot) : T_zZ \to [0, \infty]$ is convex and lower-semicontinuous,
		\item $\Psi(z, 0) = 0$,
		\item $\Psi(z, v) = \Psi(z, -v)$;
	\end{enumerate}
	the last property, called \emph{symmetry}, is sometimes replaced by the weaker condition $$\argmin\limits_{v \in T_zZ} \Psi(z, v) = 0 \, ,$$ which plays the role of an integrability condition \cite{MPR14}.
\end{itemize}

A gradient structure uniquely induces the differential equation~\eqref{GF1} and, conversely, we say that a differential equation on the space $Z$ has a gradient structure if it may be put in the form~\eqref{GF1} for at least one pair of functions~$(E, \Psi)$. Note that the latter characterization is not unique, since -- even for a fixed driving function~$E$ -- many dissipation potentials~$\Psi$ lead to the same differential equation \cite{aM19,MMP21}. Hence, assigning a gradient system to a differential equation means identifying thermodynamic information that is not available in the differential equation itself.

The derivative operators $\di$ and $\partial$ in \Eq\eqref{GF1} are abstract differential operators that, in specific cases, assume concrete representations. In a linear finite-dimensional space, like for the RDME, they become the usual gradient and the gradient with respect to a subset of the variables. In infinite-dimensional spaces, as for 
the RDPDE, they may be though of as total and partial functional derivatives.

The properties of the dissipation potential allow us to introduce a related function, called a \emph{dual dissipation potential}, as the Legendre-Fenchel transform
\begin{equation}
	\Psi^*(z, \xi) = \sup\limits_{v \in T_zZ} \left[ \pair{\xi, v}_Z - \Psi(z, v) \right] ,
\end{equation}
where $\pair{\,,}_Z : T_zZ \times T^*_zZ \to \R$ is a dual pairing between forces~$\xi$ and rates~$v$. The dual dissipation potential enjoys the same properties~1-3 above and allows us to express the gradient-flow equation~\eqref{GF1} in a different form. We apply the property of the Legendre-Fenchel transform
\begin{equation}
	\partial_\xi\Psi^*\bigl(z, \partial_v\Psi(z, v)\bigr) = v
\end{equation}
to \Eq\eqref{GF1}, obtaining
\begin{equation}\label{GF2}
	\dot{z}_t = \partial_\xi\Psi^*\bigl(z_t, - \di E(z_t)\bigr) \, .
\end{equation}

It is normally much easier, and often the only possibility, to have an explicit form for the dual dissipation potential~$\Psi^*$ rather than for the (primal) potential~$\Psi$ (cf.~for instance \cite{MPPR17}), and the same holds for the gradient-flow equation~\eqref{GF2} compared to \eqref{GF1}. This work will be no exception.
%  , and the typical dual dissipation potentials will be the quadratic form
%  \begin{equation}
	%   \Psi^*(z, \xi) = \frac{1}{2} \xi \cdot M(z) \xi \, ,
	%  \end{equation}
%  where $M(x) : T^*_zZ \to T_zZ$ is a linear operator, and the nonlinear
%  \begin{equation}
	%   \Psi^*(z, \xi) = A(z) \C(\Gamma \xi) \, ,
	%  \end{equation}
A sufficient condition for assigning a gradient structure to the RDME~\eqref{RDME} and the RDPDE~\eqref{RDPDE} is detailed balance, which we explore in the next section.

\subsection{Detailed balance} \label{sec:DB}
The RDME~\eqref{RDME} and the RDPDE~\eqref{RDPDE} have many stationary solutions, depending on the initial condition. There are, in fact, conserved quantities that force the state of the system to evolve within certain \emph{stoichiometric simplices} \cite{AK11}. General stationary states are solutions $\overline{\bbu}$ and $\overline{\bbc}$ of the equations
\begin{equation}
	\sum\limits_{\ii \in \Z_N^d} \sum\limits_{r=1}^R \mathfrak{R}^r_\ii(\overline{\bbu}) + \sum\limits_{\ii \in \Z_N^d} \sum\limits_{s=1}^S \sum\limits_{\ell=1}^d \mathfrak{D}^{s, \eell}_\ii(\overline{\bbu}) = \bbzero
\end{equation}
and
\begin{equation}
	\sum\limits_{r=1}^R \mathfrak{r}^r(\overline{\bbc}) (\bsbeta_r - \bsalpha_r) - \Div \mathfrak{d}(\overline{\bbc}) = \bbzero \, ,
\end{equation}
which result from setting \eqref{RDME} and \eqref{RDPDE} to zero, respectively. 

The detailed-balanced states constitute a subset of the class of steady states. We begin with the deterministic system and say that it satisfies \emph{deterministic detailed balance} (DDB) if there exists a state $\overline{\bbc}$ such that
\begin{subequations}
	\begin{align}
		\mathfrak{r}^r(\overline{\bbc}) 	&= 0 \quad \text{for all } r \, , \\
		\mathfrak{d}(\overline{\bbc}) 	&= \bszero \, ,
	\end{align}
\end{subequations}
that is, for every $\xx \in T^d$,
\begin{subequations}\label{DDB}
	\begin{align}
		k_{+r} \, \overline{\bsc}(\xx)^{\bsalpha_r} &= k_{-r} \, \overline{\bsc}(\xx)^{\bsbeta_r} \eqqcolon \kappa_r(\xx) \quad \text{for all } r \, , \label{RDDB} \\
		\bsD(\xx) \, \nnabla \overline{\bsc}(\xx) 	&= \bszero \, . \label{DDDB}
	\end{align}
\end{subequations}
The existence of a DDB state depends on the reaction rate constants $k_{+r}$ and $k_{-r}$ and is characterized, for the reaction part in a well-mixed system, in \cite[\Sec2.2]{MM20}.

The notion of \emph{stochastic detailed balance} (SDB) for the RDME follows the usual definition of detailed balance for a Markov process: there exists $\overline{\bbu} \in \cP(\cN)$ such that the jump rates between any two states are equilibrated. This means that, for all $\bbn$ and $\ii$,
\begin{subequations}\label{SDB}
	\begin{align}
		\overline{u}_{\bbn + \bbalpha^r_\ii} \, k_{+r} \, \bbB^{\bbalpha^r_\ii}\!(\bbn + \bbalpha^r_{\ii}) &= \overline{u}_{\bbn + \bbbeta^r_\ii} \, k_{-r} \, \bbB^{\bbbeta^r_\ii}\!(\bbn + \bbbeta^r_{\ii}) \eqqcolon \nu^{\bbn, r}_\ii && \text{for every } r , \label{RSDB} \\
		\overline{u}_{\bbn + \bbe^s_\ii} D^{s,+\eell}_\ii \, \bbB^{\bbe^s_\ii}\!(\bbn + \bbe^s_{\ii}) &= \overline{u}_{\bbn + \bbe^s_{\ii+\eell}} D^{s,-\eell}_{\ii+\eell} \, \bbB^{\bbe^s_{\ii+\eell}}(\bbn + \bbe^s_{\ii + \eell}) \eqqcolon \nu^{\bbn, s, \eell}_\ii && \text{for every } s, \eell . \label{DSDB}%,
	\end{align}
\end{subequations}

The relationship between the two notions of detailed balance is studied generally in \cite{bJ15} for purely reactive systems. If a reaction network satisfies DDB with respect to the concentration~$\overline{\bbc}$, then  the corresponding Markov process satisfies SDB with respect to the Poisson-like distributions \cite[Theorem~3.1]{MM20}
\begin{align}
	\overline{\bbu} = \chi^{\overline{\bbc}} \qquad \text{with} \qquad & \chi^\bbc(\bbn) \coloneqq e^{- \abs{\bbc}} \frac{\bbc^\bbn}{\bbn!} = \prod\limits_{\ii \in \Z_N^d} e^{- \abs{\bsc(\ii/N)}} \frac{\bsc(\ii/N)^{\bsn_\ii}}{\bsn_\ii!} \, , \label{Poisson} \\
	& \abs{\bbc} \coloneqq \sum\limits_{\ii \in \Z_N^d} \abs{\bsc(\ii/N)} \, , \quad \text{and} \quad \bbc^\bbn \coloneqq \prod\limits_{\ii \in \Z_N^d} \bsc(\ii/N)^{\bsn_\ii} \, . \nonumber
\end{align}

We may see this directly by elaborating both sides of \Eq\eqref{RSDB} when $\bsn_{\ii} \in \N^S$ (otherwise both terms vanish):
\begin{align*}
	\overline{u}_{\bbn + \bbalpha^r_\ii} \, k_{+r} \, \bbB^{\bbalpha^r_\ii}\!(\bbn + \bbalpha^r_{\ii}) &= e^{- \abs{\overline{\bbc}}} \frac{\overline{\bbc}^{\bbn+\bbalpha^r_\ii}}{(\bbn+\bbalpha^r_\ii)!} \, k_{+r} \frac{(\bbn+\bbalpha^r_\ii)!}{\bbn!} = \overline{\bbu}_\bbn \, k_{+r} \, \overline{\bsc}(\ii/N)^{\bsalpha_r} \, , \\
	\overline{u}_{\bbn + \bbbeta^r_\ii} \, k_{-r} \, \bbB^{\bbbeta^r_\ii}\!(\bbn + \bbbeta^r_{\ii}) &= e^{- \abs{\overline{\bbc}}} \frac{\overline{\bbc}^{\bbn+\bbbeta^r_\ii}}{(\bbn+\bbbeta^r_\ii)!} \, k_{-r} \frac{(\bbn+\bbbeta^r_\ii)!}{\bbn!} = \overline{\bbu}_\bbn \, k_{-r} \, \overline{\bsc}(\ii/N)^{\bsbeta_r} \, .
\end{align*}
If \Eq\eqref{RDDB} is satisfied, the two expressions are both equal to $\overline{\bbu}_\bbn \kappa_r(\ii/N)$, and we conclude that $\nu^{\bbn, r}_\ii = \overline{\bbu}_\bbn \kappa_r(\ii/N)$.

We now turn to the diffusion sector. First, we assume that the diffusion rates for the forward and backward paths between two lattice sites are equal to each other:
\begin{equation}\label{D}
	D^{s,+\eell}_\ii = D^{s,-\eell}_{\ii+\eell} \, .
\end{equation}
If we did not make this assumption, we would expect that, in the limit $N \to \infty$, diffusion would be overcome by transport governed by the difference $D^{s,+\eell} - D^{s,-\eell}$ \cite[Introduction]{KL99}: diffusion is a lower-order effect with respect to transport, and the $N^2$-scaling should be replaced by an $N$-scaling. As a consequence of \eqref{D}, there is one diffusion rate for each direction~$\ell$ and each point~$\ii$, i.e., we have a diagonal diffusion tensor. In other words, the $d$ different directions~$\ell$ are the principal directions of diffusion. We thus define, in the (principal) basis $\{\eell\}_{\ell=1,\ldots,d}$,
\begin{equation}
	D^s_{\ell\ell}(\xx) \coloneqq D^{s,+\eell}_{N \xx} = D^{s,-\eell}_{N \xx + \eell} \qquad \text{for } \xx \in T^d \, .
\end{equation}
Following these observations, and assuming \eqref{Poisson}, let us develop both sides of \Eq\eqref{DSDB}:
\begin{align*}
	\overline{u}_{\bbn + \bbe^s_{N \xx}} \, D^{s,+\eell}_{N \xx} \, \bbB^{\bbe^s_{N \xx}}(\bbn + \bbe^s_{N \xx}) &= \overline{u}_{\bbn} \, D^s_{\ell\ell}(\xx) \, \overline{c}^s(\xx) \, , \\
	\overline{u}_{\bbn + \bbe^s_{N\xx+\eell}} \, D^{s,-\eell}_{N \xx} \, \bbB^{\bbe^s_{N \xx + \eell}}(\bbn + \bbe^s_{N \xx + \eell}) &= \overline{u}_{\bbn} \, D^s_{\ell\ell}(\xx) \, \overline{c}^s(\xx + \eell/N) \, .
\end{align*}
By \Eq\eqref{DSDB} the left-hand sides agree. 
In order for the two expressions on the right-hand sides to be the same as well, we thus need, at every point~$\xx$, either a constant concentration or a vanishing diffusion coefficient. Since the diffusion tensor, in the principal basis, is diagonal, this is equivalent to say that $\bsD(\xx) \nnabla \overline{\bsc}(\xx) = 0$ for every $\xx$. We conclude that, even when reaction and diffusion are combined, DDB implies SDB with respect to Poisson-like distributions parametrized by concentrations~$\overline{c}$ that satisfy
\begin{align}
	& k_{+r} \, \overline{\bsc}(\ii/N)^{\bsalpha_r} = k_{-r} \, \overline{\bsc}(\ii/N)^{\bsbeta_r} && \text{for every $r$, and} \\
	& D^s_{\ell\ell}(\xx)\bigl(\overline{c}^s(\xx + \eell/N) - \overline{c}^s(\xx)\bigr) = 0 && \text{for every } s\text{, }\ell\text{, }\text{and } \xx.
\end{align}

When a detailed-balance condition for a system is satisfied, by experience we always expect to be able to assign a gradient structure to that system (cf., for instance, \cite{aM11,MM20}). This is true for both the deterministic and the stochastic reaction-diffusion systems that we consider in this work. The opposite implication does however not hold, since the detailed balance condition is in general only sufficient. Nevertheless, there is a situation where the notions of a gradient structure and of detailed balance are equivalent, namely when the gradient structure is derived from a large-deviation principle \cite{MPR14}: when the differential equation is the hydrodynamic limit of a certain stochastic system and a large-deviation principle holds, a dissipation potential may uniquely be derived from the large-deviation rate function if and only if the stochastic system satisfies detailed balance. Note, however, that detailed balance is not considered for the differential equation itself, but for the underlying stochastic process. This is exactly the case of the RDPDE, even if this is valid only formally since, to date, no rigorous proofs of the hydrodynamic limit from the RDME nor of the corresponding large-deviation principle have been established in full generality (partial results can be found in \cite{DMFL85,DMFL86,JLLV93,aP00,BL12,mM18}).

\subsection{Gradient structures for the RDPDE}
It is a well-known fact that, if a detailed-balance condition is satisfied, a well-mixed reaction-rate equation can be cast as a gradient flow (cf.~references below), and the same is true for a diffusion equation \cite{mG86,fO01}. The notion~\eqref{DDB} of DDB that we introduced above combines reaction and diffusion \cite{aM11}.

The choice of a gradient structure, however, is not unique, even with the same driving function (cf.~\cite[Example~4.3]{aM19} and \cite{MM20} for reaction rate equations). In this paper, we make a particular choice of a gradient structure. As a driving function, we select a relative entropy with respect to the stationary concentrations. For the dissipation potential, we choose a quadratic form for the diffusion sector, and a $\cosh$-type form for the reaction sector:
\begin{subequations}\label{GSRDPDE}
	\begin{align}
		\mathfrak{e}(\bbc) =& \int_{\T^d} \sum\limits_{s=1}^S \left( c^s(\xx) \log\frac{c^s(\xx)}{\overline{c}^s(\xx)} - c^s(\xx) + \overline{c}^s(\xx) \right) \di \xx \, , \label{GSRDPDEa} \\[3pt]
		\psi^*(\bbc, \bbmu) =& \int_{\T^d} \sum\limits_{r=1}^R \kappa_r(\xx) \left(\frac{\bsc(\xx)}{\overline{\bsc}(\xx)}\right)^{\!\!\!\frac{\bsalpha_r + \bsbeta_r}{2}} \!\!\C\bigl( (\bsbeta_r - \bsalpha_r) \cdot \bsmu(\xx)\bigr) \, \di \xx \nonumber \\
		&+ \frac{1}{2} \int_{\T^d} \sum\limits_{s=1}^S c^s(\xx) \, \bigl(\nnabla\mu^s(\xx)\bigr)^\intercal \, D^s(\xx) \, \nnabla\mu^s(\xx) \, \di \xx  \, , \label{dissipationRDPDE}
	\end{align}
\end{subequations}
where
\begin{equation} \label{eq:C}
	\C(\zeta) \coloneqq 4 \left( \cosh\frac{\zeta}{2} - 1 \right) ,
\end{equation}
\begin{equation} \label{m/n}
	\frac{\bsm}{\bsn} \coloneqq \left( \frac{m^s}{n^s}\right)_{s=1,...,S} \, ,
\end{equation}
and
\begin{equation}
	\bsm \cdot \bsn \coloneqq \sum\limits_{s=1}^S m^s \, n^s \, 
\end{equation}
for $\bsm,\bsn \in \mathbb{R}^S$. 
We deliberately avoid specifying the spaces, but only write symbolically that $\bbc \in \cC$, $\bbs \in T_\bbc\cC$, $\bbmu \in T^*_\bbc \cC$, and $\pair{\bbmu, \bbs}_{\cC} \coloneqq \int_{\T^d} \bsmu(\xx) \cdot \bss(\xx) \, \di\xx$. A possible choice for the state space, in the case we expect classical solutions of the RDPDE, is $\cC = C^2(\T^d, \R_+^S)$, but this surely does not cover the general situation, where a weaker notion of a solution is certainly needed \cite{mP10}. If we think of the limit from the RDME, which evolves measures, the most natural choice would be $\cC = \mathcal{M}_+(\T^d)^S$, the space of positive, finite vector measures on $\T^d$, which, however, brings difficulties in the definition of the gradient structure and in the interpretation of the RDPDE itself. We suggest the review paper \cite{mP10} on the well-posedness of RDPDEs for a discussion of these issues from the purely deterministic side.

The $\cosh$-type structure is advocated in \cite{mG93} and motivated by a connection between gradient flows and large deviations \cite{MPR14,MPPR17}, in contrast to the quadratic dissipation potential that is studied in \cite{OG97,aM11,MM20}. One may verify (\appendixname~\ref{App:RDPDE}) that the associated gradient flow
\begin{equation}\label{RDPDE_gradientFlow}
	\dot{\bbc}_t = \frac{\delta \psi^*}{\delta \bbmu}\biggl(\bbc_t, -\frac{\delta \mathfrak{e}}{\delta \bbc}(\bbc_t)\biggl)
\end{equation}
is indeed the RDPDE~\eqref{RDPDE}. Note that the abstract derivatives have taken the concrete expressions of functional derivatives.

%  The vector fields $\mathfrak{r}^r$ and $\mathfrak{d}$ may also be rewritten in the symmetrically balanced form \AM{Perhaps remove}
%  \begin{equation}
	%   \mathfrak{r}^r(\bbc)(\xx) = \kappa_r(\xx) \left( \frac{\bsc(\xx)^{\bsalpha_r}}{\overline{\bsc}(\xx)^{\bsalpha_r}} - \frac{\bsc(\xx)^{\bsbeta_r}}{\overline{\bsc}(\xx)^{\bsbeta_r}} \right) \qquad \text{and} \qquad \mathfrak{d}(\bbc)(\xx) = - \bsD(\xx) \nnabla \frac{\bsc(\xx)}{\overline{\bsc}(\xx)} \, .
	%  \end{equation}

\subsection{Gradient structures for the RDME}
A master equation, which governs the evolution of the probabilities of all states of a Markov process, is a special case of a reaction rate equation, one where all the reactions are of first order: the probability of each state is equivalent to the concentration of a species, and there is one ``reaction'' for each pair of different states. The ``stoichiometric coefficients'' are unit vectors in $\cN$. This observation \cite{jM11,mE14,PRST20} leads us to formulate the gradient structure for the RDME
\begin{subequations}\label{GGS-RDME}
	\begin{align}
		&\mathfrak{E}_N(\bbu) = N^{-d} \sum\limits_{\bbn \in \cN} \left( u_{\bbn} \log\frac{u_{\bbn}}{\overline{u}_{\bbn}} - u_{\bbn} + \overline{u}_{\bbn} \right) , \label{E-RDME} \\[3pt]
		&\Psi_N^*(\bbu, \bbeta) = N^{-d} \sum\limits_{\bbn \in \cN} \sum\limits_{\ii \in \Z_N^d} \sum\limits_{r=1}^R \nu^{\bbn, r}_{\ii} \biggl( \frac{u_{\bbn + \bbalpha^r_{\ii}}}{\overline{u}_{\bbn + \bbalpha^r_{\ii}}} \frac{u_{\bbn + \bbbeta^r_{\ii}}}{\overline{u}_{\bbn + \bbbeta^r_{\ii}}} \biggr)^{\!\!\frac{1}{2}} \C\bigl(N^d(\eta_{\bbn + \bbbeta^r_{\ii}} - \eta_{\bbn + \bbalpha^r_{\ii}})\bigr) \nonumber \\
		&+ N^{2-d}  \sum\limits_{\bbn \in \cN} \sum\limits_{\ii \in \Z_N^d} \sum\limits_{s=1}^S \sum\limits_{\ell=1}^d \nu^{\bbn, s, \eell}_{\ii} \biggl( \frac{u_{\bbn + \bbe^s_{\ii}}}{\overline{u}_{\bbn + \bbe^s_{\ii}}} \frac{u_{\bbn + \bbe^s_{\ii + \eell}}}{\overline{u}_{\bbn + \bbe^s_{\ii + \eell}}} \biggr)^{\!\!\frac{1}{2}} \C\bigl(N^d(\eta_{\bbn + \bbe^s_{\ii + \eell}} - \eta_{\bbn + \bbe^s_{\ii}})\bigr) \, , \label{dissipationRDME}
	\end{align}
\end{subequations}
where $\bbu \in \cP(\cN)$ and $\bbeta \in T^*_\bbu\cP(\cN)$. The tangent and cotangent spaces to $\cP(\cN)$, since $\cN$ is countable, can be both identified with the space $\cT(\cN) \coloneqq \Bigl\{ \bbeta \in \R^\cN \Bigm\vert \sum\limits_{\bbn \in \cN} \eta_\bbn = 0 \Bigr\}$ with the pairing $\pair{\bbeta, \bbv}_{\cP(\cN)} \coloneqq \sum\limits_{\bbn \in \cN} \eta_\bbn \bbv_\bbn$.

Again, we verify in \appendixname~\ref{App:RDME} that this pair of function generates the RDME~\eqref{RDME} in the gradient-flow form
\begin{equation}\label{RDME_GradientFlow}
	\dot{\bbu}_t = \frac{\partial \Psi_N^*}{\partial \bbeta}\biggl(\bbu_t, -\frac{\partial \mathfrak{E}_N}{\partial \bbu}(\bbu_t)\biggr) \, .
\end{equation}
In this finite-dimensional situation, the abstract derivatives become gradients, which we denote as partial derivatives with respect to vectors.

%  The operators $\mathfrak{R}^r_\ii$ and $\mathfrak{D}^{s, \eell}_\ii$ may also be rewritten in the symmetric form \cite{MM20} \AM{perhaps remove}
%  \begin{align}
	%   \mathfrak{R}^r_\ii(\bbu) &= \sum\limits_{\bbn \in \cN} \nu^{\bbn, r}_\ii \left( \frac{u_{\bbn + \bbalpha^r_\ii}}{\overline{u}_{\bbn + \bbalpha^r_\ii}} - \frac{u_{\bbn + \bbbeta^r_\ii}}{\overline{u}_{\bbn + \bbbeta^r_\ii}} \right) \left( \ee^{(\bbn + \bbbeta^r_\ii)} - \ee^{(\bbn + \bbalpha^r_\ii)} \right) , \nonumber \\
	%   \mathfrak{D}^{s, \eell}_\ii(\bbu) &= \sum\limits_{\bbn \in \cN} \nu^{\bbn, r}_\ii \bigg( \frac{u_{\bbn + \bbe^s_\ii}}{\overline{u}_{\bbn + \bbe^s_\ii}} - \frac{u_{\bbn + \bbe^s_{\ii+\eell}}}{\overline{u}_{\bbn + \bbe^s_{\ii+\eell}}} \bigg) \left( \ee^{(\bbn + \bbe^s_{\ii+\eell})} - \ee^{(\bbn + \bbe^s_\ii)} \right) ,
	%  \end{align}
%  where $\ee^{(\bbm)}$ is the unit vector in the direction $\bbm$, namely,
%  \begin{equation*}
	%   \ee^{(\bbm)}_\bbn = \delta_{\bbn - \bbm} \, .
	%  \end{equation*}

% \subsubsection*{Summary of the notation}
%  $\R_+$, $T_zZ$, $T^*_zZ$, $TZ$, $T^*Z$, $\di$, $\Z_N^d$, $\T^d$, introduce $\T_N^d$, $\cN \coloneqq \N^{S N^d}$, $\sym{\R^{d \times d}}$, $\C(\zeta) \coloneqq 4 \left( \cosh\frac{\zeta}{2} - 1 \right)$, $\cdot$, $\bcdot$, $\pair{}_{\cC(\T^d)}$, $\pair{}_{\cP(\cN)}$

\section{Hydrodynamic limit using gradient structures}\label{sec:limit}
In the Introduction, we showed that the hydrodynamic limit of the RDME to the RDPDE has been studied by various authors at different levels of complication. To our knowledge, the most recent work in this direction is \cite{mM18}, which, however, cannot deal with the reaction-diffusion system itself, but with a reaction-\emph{exclusion} process, where the number of particles on each lattice site is bounded from above. The \emph{a priori} locally unbounded concentrations in the deterministic reaction-diffusion system create technical difficulties that have still to be overcome.

In this paper, we consider the general case of a RDME and perform the hydrodynamic limit with a novel technique and no claim of rigor. The idea, which we sketch below, is based on the \emph{coarse-graining} strategy for gradient systems that Maas and Mielke developed in \cite[\Sec6.1]{MM20}. We will see how this method, in contrast to what happens in \cite{MM20} for the well-mixed scenario, is not sufficient to recover all information about the RDPDE: a further limit has to be taken. A central role in the success of the coarse-graining procedure is played by the local-equilibrium Poisson-like distributions~\eqref{Poisson}.

\subsection{The coarse-graining method of Maas and Mielke}
Let us consider a gradient system $(Z, E_Z, \Psi_Z)$ and its associated gradient flow
\begin{equation}
	\dot{z}_t = \partial_\xi\Psi_Z^*\bigl(z_t, -\di E_Z(z_t)\bigr) \, .
\end{equation}
Let us suppose that the solutions, for sufficiently large times, approach an exact or approximate invariant manifold $M$. On that manifold, we expect to be able to describe the system with a simplified set of variables~$y$. We thus introduce another manifold $Y$ and embed it into $Z$,
\begin{equation}
	\iota : Y \hookrightarrow Z \, ,
\end{equation}
in such a way that $\iota(Y)$ approximates $M$. The embedding~$\iota$ is called a \emph{reconstruction mapping} by Maas and Mielke, since it is the opposite of a coarse-graining map: to every coarse-grained state~$y$, it associates a fine-grained state~$z$.

In the approach of \cite{MM20}, the gradient system $(Z, E_Z, \Psi_Z)$ is pulled back by the reconstruction mapping $\iota$ to obtain the gradient system $(Y, E_Y, \Psi_Y)$:
\begin{subequations}\label{pullback}
	\begin{align}
		E_Y(y) 		&\coloneqq E_Z\bigl(\iota(y)\bigr) \, , \\
		\Psi_Y(y, s)	&\coloneqq \Psi_Z\bigl(\iota(y), \di \iota(y)(s)\bigr) \, . \label{pullbackD}
	\end{align}
\end{subequations}
We expect that the gradient flow associated with the gradient system $(Y, E_Y, \Psi_Y)$ approximates the gradient flow of $(Z, E_Z, \Psi_Z)$ in the vicinity of the invariant manifold $M$. The key step in this procedure is to find a reconstruction mapping~$\iota$ that allows us to perform this approximation successfully. The choice usually requires a great deal of experience on the system. For instance, in \cite{MM20}, where the authors consider the coarse-graining passage from the chemical master equation to the corresponding reaction rate equation, the knowledge of the stationary states is crucial.

Even once we have chosen a reconstruction mapping, the definitions~\eqref{pullback} cannot generally be used in practice, since the primal dissipation potential~$\Psi$ is usually not known explicitly, whereas the dual dissipation potential is. Maas and Mielke show how the pullback~\eqref{pullbackD} may be expressed in the dual formulation
\begin{equation}\label{min}
	\Psi^*_Y(y, \mu) = \inf\limits_{\eta \in T^*_{\iota(y)}Z}\bigl\{ \Psi^*_Z(\iota(y), \eta) \bigm\vert \di \iota(y)^\intercal(\eta) = \mu \bigr\} \, ,
\end{equation}
where $\di \iota(y)^\intercal : T^*_{\iota(y)}Z \to T^*_yY$ is the transpose map of $\di \iota(y) : T_yY \to T_{\iota(y)}Z$, namely a map that satisfies
\begin{equation}
	\left\langle \di\iota^\intercal(y)(\eta), s \right\rangle_Z = \left\langle \eta, \di\iota(y)(s) \right\rangle_Y \quad \text{for all } s \in T_yY \text{ and } \eta \in T^*_{\iota(y)}Z \, .
\end{equation}
Normally, the minimization~\eqref{min} is extremely complicated and has to be performed approximately, namely one searches for approximate minimizers $\underline{\eta} = \tilde{m}(y, \mu)$, sometimes even in the linear form $\underline{\eta} = \hat{m}(y) \mu$. An approximation may indeed be sufficient, since the whole procedure is only an approximation.

\subsection{Application to the passage from the RDME to the RDPDE}

\begin{figure}
	\centering
	\begin{tikzpicture}
		\pgfplotsset{typeset ticklabels with strut}
		\begin{axis}[width=175pt, height=175pt, domain=0:4, xmin=-.2, xmax=4.2, ymin=-.2, ymax=4.2, xtick={0,1,...,4}, ytick={0,1,...,4}, grid=both, axis line style={draw=none}, axis equal, ticks=none, font=\small, clip=false]
			\addplot[only marks, scatter, scatter src={mod(\coordindex, 3)}, samples=50] {2*rand+2};
			\draw[decoration={brace, raise=19pt}, decorate] (axis cs: .5, 0) -- node[left=19pt] {$\frac{1}{N}$} (axis cs: .5, 1);
		\end{axis}
		\begin{scope}[xshift=140]
			\draw[<->, very thin] (0, 0.15) -- (0, 3.55) node[left, pos=.5] {$1$};
		\end{scope}
		\begin{scope}[xshift=180pt]
			\begin{axis}[width=175pt, height=175pt, domain=0:4, xmin=-.2, xmax=4.2, ymin=-.2, ymax=4.2, xtick={0,0.25,...,4}, ytick={0,0.25,...,4}, grid=both, axis line style={draw=none}, axis equal, ticks=none, font=\small, mark options={scale=.25}, clip=false]
				\addplot[only marks, scatter, scatter src={mod(\coordindex, 3)}, samples=800] {2*rand+2};
				\draw[decoration={brace, raise=19pt}, decorate] (axis cs: .5, 0) -- node[left=19pt] {$\frac{1}{N}$} (axis cs: .5, .25);
			\end{axis}
		\end{scope}
	\end{tikzpicture}
	\caption{Portions of the rescaled torus $\T_N^d \coloneqq N^{-1} \Z_N^d$ for two values of $N$. In the hydrodynamic limit, the number of lattice sites is increased and the order of magnitude of the number of particles per lattice site is kept constant.}
	\label{fig:torus}
\end{figure}
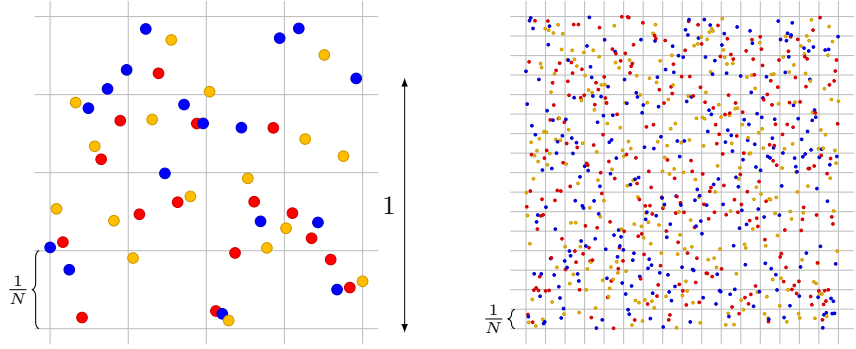

In the hydrodynamic limit of the RDME, the parameter~$N$ controls the number of lattice points. When $N$ increases, the number of particles at each point~$\ii$ remains of the same order of magnitude, which means that, globally, the total number of particles scales with~$N$ (cf.~\figurename~\ref{fig:torus}). Mathematically, this fact has to be encoded in the initial condition~$\bbu^N_0$.

We now apply the coarse-graining method to the reaction-diffusion systems, starting from the choice of a reconstruction mapping. In their work \cite{MM20}, Maas and Mielke study the passage from the chemical master equation to the reaction rate equation. When detailed balance is satisfied, the stationary states of the chemical master equation are in the form of product Poisson distributions parametrized by a positive real number that represents an equilibrium concentration. The authors then choose a reconstruction mapping that takes a generic concentration and gives a Poisson distribution parametrized by it.

Inspired by this idea, we also try the product Poisson distributions parametrized by a generic concentration, that is,
\begin{equation}\label{reconstruction}
	\functiondef{\iota_N}{\cC_N}{\cP(\cN) \, ,}{\bbc}{\chi^{\bbc}_N \, ,}
\end{equation}
where $\cC_N$ is a set of functions (or measures) on the rescaled discrete lattice $\T_N^d \coloneqq N^{-1} \Z_N^d$, and we have introduced an index $N$ to remark the $N$-dependence. Note that $(\T_N^d)_{N \in \N}$ is a sequence of sets that becomes denser and denser in $\T^d$. We indicate the tangent and cotangent spaces by $T_\bbc\cC_N$ and $T^*_\bbc\cC_N$, and their dual pairing by
\begin{equation}
	\pair{\bbmu, \bbs}_{\cC_N} \coloneqq N^{-d} \, \bbmu \bcdot \bbs \coloneqq N^{-d} \sum\limits_{\ii \in \Z^d_N} \bsmu(\ii/N) \cdot \bss(\ii/N) \, .
\end{equation}

The choice~\eqref{reconstruction} presents the same feature of entailing the correct stationary distributions~$\chi^{\overline{\bbc}}$ when $\bbc = \overline{\bbc}$, but shows an even more interesting phenomenon. In the limit $N \to \infty$, the distribution $\chi^{\bbc}$ embodies the concept of \emph{local equilibrium}: since diffusion is $N^2$ times faster than reaction (cf.~the scaling in \Eqs\eqref{reactionVectorField}-\eqref{diffusionVectorField} and \eqref{dissipationRDME}), for large $N$, diffusion locally equilibrates the particles between neighboring lattice sites. This means that, as soon as a reaction pushes the system away from a local product Poisson distribution, diffusion immediately restores that distribution locally. In other words, the space of product Poisson distributions is an attractive invariant manifold for the RDME. This phenomenon of local equilibration, for reaction-exclusion processes, is examined rigorously in the replacement lemma, \Th3.5, of \cite{mM18}.

Let us now consider the gradient structure $(\mathfrak{E}_N, \Psi_N)$ of the RDME and pull back the driving function~$\mathfrak{E}_N$ under the reconstruction mapping:
\begin{equation}\label{pullbackEntropy}
	\mathfrak{E}_N(\iota_N(\bbc)) = N^{-d} \sum\limits_{\ii \in \Z_N^d} \sum\limits_{s=1}^S \left( c^s(\ii/N) \log\frac{c^s(\ii/N)}{\overline{c}^s(\ii/N)} - c^s(\ii/N) + \overline{c}^s(\ii/N) \right) .
\end{equation}
We immediately note that, as anticipated, the parameter~$N$ has not disappeared from the expression.

To find the dual dissipation potential, we first have to search for an approximate solution of
\begin{equation}\label{inf}
	\inf\limits_{\bbeta \in T^*_{\iota_{N}(\bbc)}\cP(\cN)}\biggl\{ \Psi^*_N(\iota_N(\bbc), \bbeta) \biggm\vert \biggl(\frac{\partial \iota_N(\bbc)}{\partial \bbc}\biggr)^{\!\intercal} \, \bbeta = \bbmu \biggr\} .
\end{equation}
We thus need the derivative $\partial \iota_N(\bbc) / \partial\bbc : T_\bbc \cC_N\to T_{\iota_N(\bbc)}\cP(\cN)$ and its transpose map $(\partial\iota_N(\bbc) / \partial\bbc)^\intercal : T^*_{\iota_N(\bbc)}\cP(\cN) \to T^*_\bbc \cC_N$:
\begin{align}
	\biggl(\frac{\partial\iota_N(\bbc)}{\partial\bbc} \, \bbs\biggr)_{\!\bbn} &= \chi^\bbc_N(\bbn) \left( \frac{\bbn}{\bbc} - \bbone \right) \bcdot \bbs \, , \\
	\biggl( \Bigl(\frac{\partial \iota_N(\bbc)}{\partial \bbc}\Bigr)^{\!\intercal} \, \bbeta \biggr)(\ii/N) &= N^d \sum\limits_{\bbn \in \cN} \chi^\bbc_N(\bbn) \left( \frac{\bsn_\ii}{\bsc(\ii/N)} - \bsone \right) \bbeta_\bbn \, .
\end{align}
The following calculations completely reproduce the analogous calculations in \cite[\Sec6.2]{MM20}, apart from the additional spatial dependence. We look for a minimizer of~\eqref{inf} in the linear form $\underline{\bbeta} = \hat{m}_N(\bbc) \, \bbmu$ that approximates the true minimizer as $N \to \infty$. % \AM{(ask Maas or Mielke)}. 
We then try $\underline{\bbeta}_\bbn = \bba \bcdot \bbn$ with $\bba \in T^*_\bbc\cC_N$ and obtain
\begin{align}
	&\biggl( \Bigl(\frac{\partial\iota_N(\bbc)}{\partial\bbc}\Bigr)^{\!\intercal} \, \underline{\bbeta} \biggr)^{\!\!s}(\ii/N) \nonumber \\
	&\hspace{45pt} = N^d \sum\limits_{\bbn \in \cN} \chi^\bbc_N(\bbn) \frac{n^s_\ii}{c^s(\ii/N)} \biggl( a^s(\ii/N) \, n^s_\ii + \sum\limits_{(\jj, u) \neq (\ii, s)} a^u(\jj/N) \, n^u_\jj \biggr) - N^{-d} \sum\limits_{\bbn \in \cN} \chi^\bbc_N(\bbn) \, \bba \bcdot \bbn \nonumber \\
	&\hspace{45pt} = + N^d \Bigl( a^s(\ii/N) \bigl( c^s(\ii/N) + 1 \bigr) + \sum\limits_{(\jj, u) \neq (\ii, s)} a^u(\jj/N) \, c^u(\jj/N) \Bigr) - N^d \sum\limits_\jj \bsa(\jj/N) \cdot \bsc(\jj/N) \nonumber \\
	&\hspace{45pt} = N^d \, a^s(\ii/N) \, .
\end{align}
Hence, we conclude that
\begin{equation}
	\bigl(\hat{m}_N(\bbc) \, \bbmu\bigr)_\bbn = N^{-d} \left( \bbmu \bcdot \bbn \right)
\end{equation}
and, inserting this result into~\eqref{inf},
\begin{align}\label{pullbackDissipation}
	& \Psi_N^*(\iota_N(\bbc), \hat{m}_N(\bbc) \, \bbmu) = N^{-d} \sum\limits_{\ii \in \Z_N^d} \sum\limits_{r=1}^R \kappa^r(\ii/N) \left(\frac{\bsc(\ii/N)}{\overline{\bsc}(\ii/N)}\right)^{\!\!\!\frac{\bsalpha_r + \bsbeta_r}{2}} \!\! \C\bigl((\bsbeta_r - \bsalpha_r) \cdot \bsmu(\ii/N)\bigr) \\
	&\hspace{12pt} + N^{2-d} \sum\limits_{\ii \in \Z_N^d} \sum\limits_{s=1}^S \sum\limits_{\ell=1}^d D^s_{\ell\ell}(\ii/N) \, \overline{c}^s(\ii/N) \left( \frac{c^s(\ii/N)}{\overline{c}^s(\ii/N)} \frac{c^s((\ii+\eell)/N)}{\overline{c}^s((\ii+\eell)/N)} \right)^{\!\!\frac{1}{2}} \C\!\big(\mu^s((\ii+\eell)/N) - \mu^s(\ii/N)\big) \, . \nonumber
\end{align}
We note, as expected, that the macroscopic reaction-rate factors are the expected values of the microscopic ones with respect to the local Poisson distributions, as the following calculation demonstrates:
\begin{equation}
	\bbb^{\bbgamma}(\bbc) = \mathbb{E}_{\chi^{\bbc}}[\bbB^{\bbgamma}] = \sum\limits_{\bbn \in \cN} e^{-\abs{\bbc}} \frac{\bbc^{\bbn}}{\bbn!} \frac{\bbn!}{(\bbn - \bbgamma)!} = \sum\limits_{\bbn \in \cN} e^{-\abs{\bbc}} \frac{\bbc^{\bbn - \bbgamma}}{(\bbn - \bbgamma)!} \bbc^{\bbgamma} = \bbc^{\bbgamma} \, .
\end{equation}

\subsection{The limit $N \to \infty$}
As we had announced, the pair given by \eqref{pullbackEntropy} and \eqref{pullbackDissipation} does not give us the gradient structure~\eqref{GSRDPDE} associated with the RDPDE: the expressions still contain the discreteness of the lattice. The above procedure, nevertheless, has produced something useful once it is combined with the limit $N \to \infty$. The limit of the driving function follows immediately from the definition of a definite integral:
\begin{equation}
	\lim\limits_{N \to \infty} \mathfrak{E}_N\bigl(\iota_N(\bbc)\bigr) = \mathfrak{e}(\bbc) \, 
\end{equation}
with $\mathfrak{e}$ given in \eqref{GSRDPDEa}.

The analogous limit for the dissipation potential requires some more case. The difference in the $N$-scaling in the two factors (reaction and diffusion) plays a central role. While we can carry out the limit immediately in the reaction part, in the diffusion one we calculate the Maclaurin expansion
\begin{equation}
	N^2 \left( \frac{c^s(\ii/N)}{\overline{c}^s(\ii/N)} \frac{c^s((\ii+\eell)/N)}{\overline{c}^s((\ii+\eell)/N)} \right)^{\!\!\frac{1}{2}} \C\bigl(\mu^s((\ii+\eell)/N) - \mu^s(\ii/N)\bigr)	= \frac{c^s(\ii/N)}{\overline{c}^s(\ii/N)} \, \bigl\lVert\partial_\ell\mu^s(\ii/N)\bigr\rVert^2 + o(1) \, .
\end{equation}
By performing the summations, we thus obtain the dissipation potential~\eqref{dissipationRDPDE} with a diagonal diffusion tensor, namely
\begin{equation}
	\lim\limits_{N \to \infty} \Psi_N^*\bigl(\iota_N(\bbc), \hat{m}_N(\bbc) \, \bbmu\bigr) = \psi(\bbc, \bbmu) \, .
\end{equation}

While the coarse-graining procedure could not produce the RDPDE from the RDME directly, it still proved a valuable tool: its combination with the limit $N \to \infty$ gives us the correct answer, the expected gradient structure for the RDPDE. The coarse-graining method appears to facilitate the limit procedure, and the choice of the reconstruction mapping, which encodes the important property of local equilibrium, is the key step to ensure the success. Two important remarks are in order at this point.

(1) At least at a heuristic level -- which is the one at which we conducted all calculations and ``proofs'' -- using a gradient structure to perform the hydrodynamic limit is not necessary. A weaker \emph{variational structure} would be enough, namely a functional that is minimized at the solutions of the RDME and is in the form
\begin{equation}
	\cI^N\bigl(\bbu_t\rvert_{[t \in [0, T]}\bigr) = \cI_0^N(\bbu_0) + \int_0^T \cL(\bbu_t, \dot{\bbu}_t) \, \di t \, .
\end{equation}
The dissipation potential is replaced by the \emph{Lagrangian}~$\cL$ and the dual dissipation potential by the \emph{Hamiltonian}~$\cH$, the Legendre-Fenchel transform of~$\cL$. The use of variational structures does not require, in principle, the assumption of detailed balance, although its absence would likely complicate the rigorous proofs.

(2) We have seen that the coarse-graining method of Maas and Mielke, in the hydrodynamic limit of reaction-diffusion systems, has to be supplied with a limit procedure. Coarse-graining and limit, indeed, are two ingredients of the same recipe and fit well together in a variational framework \cite{DLPS17,HPST20}. In a future work, we will demonstrate how the variational framework of these two papers can be applied to the hydrodynamic limit of reaction-diffusion systems.

\section{Conclusions}\label{sec:conclusion}
The hydrodynamic limit of reaction-diffusion master equations to corresponding PDEs is rather clear from the formal standpoint, but a rigorous proof is still lacking. The aim of the present work is \emph{not} to fill this gap, but to suggest an alternative route through the use and strengthening of a coarse-graining strategy proposed by Maas and Mielke in \cite{MM20}. This path may represent, on the one hand, the first step towards a rigorous proof and, on the other hand, a valuable tool for the scaling limits of different systems in physical, biological, and social sciences.

Let us review our results in more detail. In a first step, we formulate the two models in a common language. The reaction-diffusion master equation (RDME) describes the stochastic dynamics of particle numbers on a $d$-dimensional lattice with $N^d$ sites. The reaction-diffusion PDE (RDPDE) deals with concentrations that evolve deterministically on a $d$-dimensional continuous space. Our notation clarifies how, at the stochastic level, reaction and diffusion may be both described as reactions in a suitably abstract sense: reaction occurs at the same lattice site, whereas diffusion involves neighboring sites. At the deterministic level, conversely, reaction and diffusion are qualitatively different.

If one looks more closely, the difference emerges already quantitatively at the stochastic level once we vary the number of lattice sites: reaction and diffusion are indeed scaled differently with respect to~$N$, and in particular diffusion happens at rates that are $N^2$ times faster than reaction. This means that, at small spatial scales and for large $N$, reaction loses the competition with diffusion, which tends to restore local-equilibrium distributions very quickly. These local-equilibrium distributions are in a Poisson-like form and play a central role in this paper.

The second step of this work is to introduce gradient-flow structures for the RDME and the RDPDE in the case where both satisfy detailed balance. Such structures are defined in terms of a driving function and a (dual) dissipation potential. The driving function is in a relative-entropy form. For the dissipation potential, although there exist multiple choices for both models, we select $\cosh$-type functional forms, which are compatible with the large deviations of the RDME according to the connection between large deviations and gradient flows that was proven in \cite{MPR14}. Any other gradient structure, however, would work as well.

Our last, main step is to study the limit passage from the RDME to the RDPDE through the coarse-graining method for gradient structures introduced in \cite{MM20}. Instead of studying the limit directly, the aim is to infer, by a simplification argument, the driving function and the dissipation potential for the RDPDE from the corresponding objects associated with the RDME. The procedure starts by introducing a map, called \emph{reconstruction mapping}, that takes a coarse-grained state (a concentration for the RDPDE model) and gives a fine-grained one (a probability measure for the RDME model). A clever choice for this map is the local-equilibrium Poisson-like distribution.

Once the choice has been made, we perform two operations. First, we pull back the driving function of the RDME by the reconstruction mapping. Second, we approximately solve a constrained minimization for the dissipation potential of the RDME on the same lines of \cite[\Sec6.2]{MM20}. In contrast to this reference, the reduced objects are not yet the driving function and dissipation potential of the RDPDE, which can only be recovered after performing a further limit. We remark, however, that this limit is much simpler than the direct limit of the initial gradient-flow structure, which may be performed, for instance, by EDP-convergence techniques \cite{SS04,MMP21}. The ``coarse-graining'' step has greatly simplified the limit procedure.

Hence, in the application presented in this paper, we give more light to the coarse-graining strategy proposed by Maas and Mielke by disentangling the ``simplification'' passage from the ``limit'' one. The simplification method, indeed, does not produce, in general, the gradient-flow structure of the coarse-grained system directly. It does, however, constitute a great advantage when we wish to take a limit. The reason for the success lies in the choice of the reconstruction mapping, which encodes the key information of local-equilibrium. This property interacts nicely with the difference in the $N$-scaling in reaction and diffusion, leading to the correct gradient-flow structures in the limit. No such a phenomenon is seen in \cite[\Sec6]{MM20}.

Although this work does not provide a rigorous proof of the limit RDME $\to$ RDPDE, it shows an example where a coarse-graining step may formally simplify the limit passage. This brings the method of \cite{MM20} closer to similar methods, such as the variational approach to coarse-graining presented in \cite{DLPS17}, which show both benefits of being rigorous and of not being restricted to detailed-balanced systems. Exploring the relation between the methods is certainly one of our next major goals, one that would help us to construct a clearer picture on the available methods for the simplification of complex systems, especially with regard to a wide range of applications in physics (chemical reaction systems), biology (signaling processes), social science (agent-based modeling).

\paragraph{Acknowledgments.}
 The research of AM was funded by the Swiss National Science Foundation via the Early Postdoc.Mobility fellowship. The research of CS and SW was partially funded by the Deutsche Forschungsgemeinschaft (DFG, German Research Foundation) under Germany’s Excellence Strategy -- The Berlin Mathematics Research Center MATH+ (EXC-2046/1 project ID: 390685689), and via Project C03 of CRC 1114.

\sloppy
\bibliography{Bibliography}

\appendix

\small

\section{Gradient flows for the RDPDE and the RDME}\label{app}

\subsection{RDME} \label{App:RDME}

We show that the gradient-flow
\[  \dot{\bbu}_t = \frac{\partial \Psi_N^*}{\partial \bbeta}\biggl(\bbu_t, -\frac{\partial \mathfrak{E}_N}{\partial \bbu}(\bbu_t)\biggr)
\]
given in \eqref{RDME_GradientFlow} generates the RDME~\eqref{RDME}. 

At first, we note that the gradient of the driving function $\mathfrak{E}_N$ defined in \eqref{E-RDME} is given component-wise by
\begin{equation}\label{E_N}
	\left(\frac{\partial\mathfrak{E}_N}{\partial \bbu}(\bbu)\right)_{\!\!\bbm} = N^{-d} \log\frac{u_\bbm}{\overline{u}_{\bbm}} \, .
\end{equation}

To determine the partial gradient $\frac{\partial\Psi_N^*}{\partial\bbeta}$ of the dissipation potential $\Psi_N^*(\bbu,\bbeta)$ given in \eqref{dissipationRDME}, we need the derivative of $\C$ defined in \eqref{eq:C}, which is given by
\begin{equation} \label{C_derivative}
	\frac{d}{d\zeta}\C(\zeta) = 2  \sinh\frac{\zeta}{2}  = \exp\Bigl(\frac{\zeta}{2}\Bigr)-\exp\Bigl(-\frac{\zeta}{2}\Bigr) .
\end{equation}

\subsubsection*{Reaction part}

Starting with the reaction part (i.e., the first line of \eqref{dissipationRDME}) and fixing $\ii \in \Z_N^d$ and $r\in \{1,...,R\}$,  we calculate
\[ \frac{\partial}{\partial\eta_{\bbm}}\C\bigl(N^d(\eta_{\bbn + \bbbeta^r_{\ii}} - \eta_{\bbn + \bbalpha^r_{\ii}})\bigr) = 2N^d \bigl(\delta_{\bbm,\bbn + \bbbeta^r_{\ii}} - \delta_{\bbm,\bbn + \bbalpha^r_{\ii}}\bigr) \sinh\frac{N^d\bigl(\eta_{\bbn + \bbbeta^r_{\ii}} - \eta_{\bbn + \bbalpha^r_{\ii}}\bigr)}{2}\]
for any $\bbm \in \cN$, where $\delta$ is the Kronecker delta. Inserting  $-\frac{\partial\mathfrak{E}_N}{\partial \bbu}(\bbu)$ given in \eqref{E_N} for $\bbeta$, i.e., replacing $\eta_{\bbm}$ by $-N^{-d} \log\frac{u_\bbm}{\overline{u}_{\bbm}}$, we obtain
\begin{align*}
	&  2N^d \, \bigl(\delta_{\bbm,\bbn + \bbbeta^r_{\ii}} - \delta_{\bbm,\bbn + \bbalpha^r_{\ii}}\bigr) \, \sinh\biggl[-\frac{1}{2}\biggl(\log \frac{u_{\bbn + \bbbeta^r_{\ii}}}{\overline{u}_{\bbn + \bbbeta^r_{\ii}}} - \log \frac{u_{\bbn + \bbalpha^r_{\ii}}}{\overline{u}_{\bbn + \bbalpha^r_{\ii}}}\biggr)\biggr] \\
	= \;\; &  2N^d \, \bigl(\delta_{\bbm,\bbn + \bbbeta^r_{\ii}} - \delta_{\bbm,\bbn + \bbalpha^r_{\ii}}\bigr) \, \sinh\biggl[ \frac{1}{2}\log \biggl(\frac{u_{\bbn + \bbalpha^r_{\ii}}}{u_{\bbn + \bbbeta^r_{\ii}}}\frac{\overline{u}_{\bbn + \bbbeta^r_{\ii}}}{\overline{u}_{\bbn + \bbalpha^r_{\ii}}} \biggr)\biggr] \\
	\overset{\eqref{eq:C}}{=} &  N^d \, \bigl(\delta_{\bbm,\bbn + \bbbeta^r_{\ii}} - \delta_{\bbm,\bbn + \bbalpha^r_{\ii}}\bigr) \left[
	\biggl(\frac{u_{\bbn + \bbalpha^r_{\ii}}}{u_{\bbn + \bbbeta^r_{\ii}}}\frac{\overline{u}_{\bbn + \bbbeta^r_{\ii}}}{\overline{u}_{\bbn + \bbalpha^r_{\ii}}} \biggr)^{\!\!\frac{1}{2}}
	-\biggl(\frac{u_{\bbn + \bbbeta^r_{\ii}}}{u_{\bbn + \bbalpha^r_{\ii}}}\frac{\overline{u}_{\bbn + \bbalpha^r_{\ii}}}{\overline{u}_{\bbn + \bbbeta^r_{\ii}}} \biggr)^{\!\!\frac{1}{2}} 
	\right] .
\end{align*} 
With this, the contribution to the gradient flow from the reaction part is
\begin{align*}
	&\sum\limits_{\bbn \in \cN} \! \nu^{\bbn, r}_{\ii} \biggl(\! \frac{u_{\bbn + \bbalpha^r_{\ii}}}{\overline{u}_{\bbn + \bbalpha^r_{\ii}}} \frac{u_{\bbn + \bbbeta^r_{\ii}}}{\overline{u}_{\bbn + \bbbeta^r_{\ii}}} \!\biggr)^{\!\!\frac{1}{2}} \! \left[\!\biggl(\!\frac{u_{\bbn + \bbalpha^r_{\ii}}}{u_{\bbn + \bbbeta^r_{\ii}}}\frac{\overline{u}_{\bbn + \bbbeta^r_{\ii}}}{\overline{u}_{\bbn + \bbalpha^r_{\ii}}} \!\biggr)^{\!\!\frac{1}{2}} \!\!-\! \biggl(\!\frac{u_{\bbn + \bbbeta^r_{\ii}}}{u_{\bbn + \bbalpha^r_{\ii}}}\frac{\overline{u}_{\bbn + \bbalpha^r_{\ii}}}{\overline{u}_{\bbn + \bbbeta^r_{\ii}}} \!\biggr)^{\!\!\frac{1}{2}} \right] \!\bigl(\delta_{\bbm,\bbn + \bbbeta^r_{\ii}} - \delta_{\bbm,\bbn + \bbalpha^r_{\ii}}\bigr) \\
	=\;\; & \sum\limits_{\bbn \in \cN}  \nu^{\bbn, r}_{\ii}  \biggl(  \frac{u_{\bbn + \bbalpha^r_{\ii}}}{\overline{u}_{\bbn + \bbalpha^r_{\ii}}}-\frac{u_{\bbn + \bbbeta^r_{\ii}}}{\overline{u}_{\bbn + \bbbeta^r_{\ii}}}  \biggr)\bigl(\delta_{\bbm,\bbn + \bbbeta^r_{\ii}} - \delta_{\bbm,\bbn + \bbalpha^r_{\ii}}\bigr) \\
	\overset{\eqref{RSDB}}{=}  & 
	\sum\limits_{\bbn \in \cN} 
	\bigl[ k_{+r} \, \bbB^{\bbalpha^r_{\ii}}(\bbn + \bbalpha^r_{\ii}) \, u_{\bbn + \bbalpha^r_{\ii}}
	-k_{-r} \, \bbB^{\bbbeta^r_{\ii}}(\bbn + \bbbeta^r_{\ii}) \, u_{\bbn + \bbbeta^r_{\ii}}  \bigr]\bigl(\delta_{\bbm,\bbn + \bbbeta^r_{\ii}} - \delta_{\bbm,\bbn + \bbalpha^r_{\ii}}\bigr) \\
	= \;\; &   
	k_{+r} \, \bigl[ \B^{\bsalpha_r}\!(\bsn_{\ii} + \bsalpha_r - \bsbeta_r) \, u_{\bbm + \bbalpha^r_{\ii}- \bbbeta^r_{\ii}} 
	- \B^{\bsalpha_r}\!(\bsn_{\ii}) \, u_{\bbm }\bigr] + k_{-r} \, \bigl[\B^{\bsbeta_r}\!(\bsn_{\ii} - \bsalpha_r + \bsbeta_r) \, u_{\bbm + \bbbeta^r_{\ii}- \bbalpha^r_{\ii}}  
	- \B^{\bsbeta_r}\!(\bsn_{\ii}) \, u_{\bbm} \bigr] \\
	=\;\; & \bigl(\mathfrak{R}^r_\ii(\bbu)\bigr)_{\bbm} \, ,
\end{align*}
which is the vector field $\mathfrak{R}^r_\ii(\bbu)$ defined in \eqref{reactionVectorField}.

\subsubsection*{Diffusion part}

Equivalently, we obtain for the diffusion part (second line of \eqref{dissipationRDME})
\begin{align*}
	& N^2\sum\limits_{\bbn \in \cN}  
	\nu^{\bbn, s, \eell}_{\ii} \bigg( \frac{u_{\bbn + \bbe^s_{\ii}}}{\overline{u}_{\bbn + \bbe^s_{\ii}}} \frac{u_{\bbn + \bbe^s_{\ii + \eell}}}{\overline{u}_{\bbn + \bbe^s_{\ii + \eell}}} \bigg)^{\!\!\frac{1}{2}} 
	\left[\biggl(\frac{u_{\bbn + \bbe^s_{\ii}}}{u_{\bbn + \bbe^s_{\ii + \eell}}}\frac{\overline{u}_{\bbn + \bbe^s_{\ii + \eell}}}{\overline{u}_{\bbn + \bbe^s_{\ii}}} \biggr)^{\!\!\frac{1}{2}} - \biggl(\frac{u_{\bbn + \bbe^s_{\ii + \eell}}}{u_{\bbn + \bbe^s_{\ii}}}\frac{\overline{u}_{\bbn + \bbe^s_{\ii}}}{\overline{u}_{\bbn + \bbe^s_{\ii + \eell}}} \biggr)^{\!\!\frac{1}{2}}\right] \bigl(\delta_{\bbm,\bbn + \bbe^s_{\ii + \eell}} - \delta_{\bbm,\bbn + \bbe^s_{\ii}}\bigr) \\
	=\;\; &   N^2 \sum\limits_{\bbn \in \cN}  \nu^{\bbn, s, \eell}_{\ii}  \biggl( \frac{u_{\bbn + \bbe^s_{\ii}}}{\overline{u}_{\bbn + \bbe^s_{\ii}}}-\frac{u_{\bbn + \bbe^s_{\ii + \eell}}}{\overline{u}_{\bbn + \bbe^s_{\ii + \eell}}} \biggr) \bigl(\delta_{\bbm,\bbn + \bbe^s_{\ii + \eell}} - \delta_{\bbm,\bbn + \bbe^s_{\ii}}\bigr) \\
	\overset{\eqref{DSDB}}{=}  & 
	N^2 \sum\limits_{\bbn \in \cN} 
	\left[ D^{s,+\eell}_\ii \,  \bbB^{\bbe^s_\ii}(\bbn + \bbe^s_{\ii}) \, u_{\bbn + \bbe^s_{\ii}}
	-D^{s,-\eell}_{\ii+\eell} \,  \bbB^{\bbe^s_{\ii+\eell}}(\bbn + \bbe^s_{\ii + \eell}) \, u_{\bbn + \bbe^s_{\ii + \eell}} \right] \bigl(\delta_{\bbm,\bbn + \bbe^s_{\ii + \eell}} - \delta_{\bbm,\bbn + \bbe^s_{\ii}}\bigr) \\
	%= \;\; & 
	%N^2 \sum\limits_{\bbn \in \cN} 
	%\left[D^{s,+\eell}_\ii \, \B^{\bse^s}(\bsn_{\ii})u_{\bbn + \bbe^s_{\ii}}
	%-D^{s,-\eell}_{\ii+\eell} \, \B^{\bse^s}(\bsn_{\ii+\eell})u_{\bbn + \bbe^s_{\ii + \eell}}  
	%\right](\delta_{\bbm,\bbn + \bbe^s_{\ii + \eell}} - \delta_{\bbm,\bbn + \bbe^s_{\ii}}) \\
	= \; \; &  N^2 D^{s,+\eell}_\ii \bigl[ (m^s_\ii + 1) \, u_{\bbm + \bbe^s_\ii - \bbe^s_{\ii+\eell}} - m^s_\ii \, u_\bbm \bigr] \\
	& + N^2 D^{s,-\eell}_{\ii+\eell} \bigl[ (m^s_{\ii+\eell} + 1) \, u_{\bbm - \bbe^s_\ii + \bbe^s_{\ii+\eell}} - m^s_{\ii+\eell} \, u_\bbm \bigr] \\
	= \;\; & \bigl(\mathfrak{D}^{s, \eell}_\ii(\bbu)\bigr)_\bbm \, ,
\end{align*}
the vector field $\mathfrak{D}^{s, \eell}_\ii(\bbu)$ defined in \eqref{diffusionVectorField}. 
In combination with the reaction part, this delivers the RDME in the form \eqref{RDME}.

\subsection{RDPDE} \label{App:RDPDE}

We show that the gradient flow 
\begin{equation*}
	\dot{\bbc}_t = \frac{\delta \psi^*}{\delta \bbmu}\biggl(\bbc_t, -\frac{\delta \mathfrak{e}}{\delta \bbc}(\bbc_t)\biggl)
\end{equation*}
given in \eqref{RDPDE_gradientFlow}  generates the RDPDE \eqref{RDPDE}.

The functional derivative of the driving function $\mathfrak{e}$ defined in \eqref{GSRDPDEa} is given by 
\[ %(\di \mathfrak{e}(\bbc))(\xx) =
\frac{\delta \mathfrak{e}}{\delta \bbc}(\bbc)(\xx) =\log \frac{\bsc(\xx)}{\overline{\bsc}(\xx)} \, ,  \]
where definition \eqref{m/n} applies and the $\log$-function acts component-wise according to
\[ \log\bsc := (\log c^s)_{s=1,...,S} \, . \] 

\subsubsection*{Functional derivative (in general)}

Given a function $f: \mathbb{R}^3 \to \mathbb{R}$, $f(x,y,z)$, and a function $g:\mathbb{R}\to \mathbb{R}$, consider the functional $F(g)= \int f(x,g(x),\nabla g(x)) \, \di x$. Then the functional derivative of $F$ with respect to $g$ is given by
\[ \frac{\delta F}{\delta g}(g)(x) = \frac{\partial}{\partial y}f\bigl(x,g(x),\nabla g(x)\bigr) - \Div \frac{\partial }{\partial z} f\bigl(x,g(x),\nabla g(x)\bigr) \, . \]
More generally, given $\bsg: \mathbb{T}^d \to \mathbb{R}^S$, $\bsy \in \mathbb{R}^S$, $\bsz \in \mathbb{R}^{d \times S}$ and $f:\mathbb{T}^d\times \mathbb{R}^S \times \mathbb{R}^{d \times S} \to \mathbb{R}$, we have  
\[ \frac{\delta F}{\delta \bbg}(\bbg)(\xx) = \frac{\partial }{\partial \bsy} f\bigl(\xx,\bsg(\xx),\nnabla \bsg(\xx)\bigr) - \Div \frac{\partial }{\partial \bsz} f\bigl(\xx,\bsg(\xx),\nnabla \bsg(\xx)\bigr) \, ,\] 
where $\nnabla \bsg =(\nnabla g^s)_{s=1,...,S}\in \mathbb{R}^{d \times S}$ and $\Div$ was defined analogously in~\eqref{divergence}.

In the following, we write 
$\psi^*(\bbc, \bbmu) =F^{\text{reac}}_{\bbc}(\bbmu) + F^{\text{diff}}_{\bbc}(\bbmu)$
to distinguish between the reaction and diffusion part of $\psi^*(\bbc, \bbmu)$ given in \eqref{dissipationRDPDE}. 

\subsubsection*{Reaction part}

According to the first line of \eqref{dissipationRDPDE} we have $F^{\text{reac}}_{\bbc}(\bbmu)  = \int_{\mathbb{T}^d}f^{\text{reac}}_{\bbc}(\xx,\bsmu(\xx),\nnabla \bsmu (\xx))\, \di \xx$ with 
\[ f^{\text{reac}}_{\bbc}(\xx,\bsy,\bsz) = \sum\limits_{r=1}^R \kappa_r(\xx) \left(\frac{\bsc(\xx)}{\overline{\bsc}(\xx)}\right)^{\!\!\!\frac{\bsalpha_r + \bsbeta_r}{2}} \!\! \C\bigl( (\bsbeta_r - \bsalpha_r) \cdot \bsy\bigr) \, . \]
Using again Eq. \eqref{C_derivative}, we obtain
\[ \frac{\partial }{\partial \bsy}f^{\text{reac}}_{\bbc}(\xx,\bsy,\bsz) = \sum\limits_{r=1}^R \kappa_r(\xx) \left(\frac{\bsc(\xx)}{\overline{\bsc}(\xx)}\right)^{\!\!\!\frac{\bsalpha_r + \bsbeta_r}{2}} \! 2 (\bsbeta_r - \bsalpha_r) \sinh\frac{(\bsbeta_r - \bsalpha_r)\cdot \bsy}{2} \, ,  \]
while $\frac{\partial }{\partial \bsz}f^{\text{reac}}_{\bbc} =0$. 
Hence, the functional derivative $\frac{\delta F^{\text{reac}}_{\bbc}}{\delta \bbmu} $ is given by 
\[ \frac{\delta F^{\text{reac}}_{\bbc}}{\delta \bbmu}(\bbmu)(\xx) = 2 \sum\limits_{r=1}^R \kappa_r(\xx) \left(\frac{\bsc(\xx)}{\overline{\bsc}(\xx)}\right)^{\!\!\!\frac{\bsalpha_r + \bsbeta_r}{2}} \! (\bsbeta_r - \bsalpha_r) \sinh \frac{(\bsbeta_r - \bsalpha_r)\cdot \bsmu(\xx)}{2} \, .\]
Inserting  $-\frac{\delta \mathfrak{e}}{\delta \bbc}$ for $\bbmu$ 
and using the equality
\[ 2 \sinh\!\left(-\frac{\bsbeta_r - \bsalpha_r}{2}\cdot \log \frac{\bsc(\xx)}{\overline{\bsc}(\xx)}\right) =\left(\frac{\bsc(\xx)}{\overline{\bsc}(\xx)}\right)^{\!\!\!\frac{\bsalpha_r - \bsbeta_r}{2}} \! -\left(\frac{\bsc(\xx)}{\overline{\bsc}(\xx)}\right)^{\!\!\!\frac{ \bsbeta_r-\bsalpha_r}{2}} \, , \]
we get
\begin{align*}
	\frac{\delta F^{\text{reac}}_{\bbc}}{\delta \bbmu}\biggl(-\frac{\delta \mathfrak{e}}{\delta \bbc}(\bbc)\biggr)(\xx) &= \sum\limits_{r=1}^R\kappa_r(\xx) \left( \frac{\bsc(\xx)}{\overline{\bsc}(\xx)} \right)^{\!\!\!\frac{\bsalpha_r + \bsbeta_r}{2}}\!\! \left[\left(\frac{\bsc(\xx)}{\overline{\bsc}(\xx)}\right)^{\!\!\!-\frac{\bsbeta_r - \bsalpha_r}{2}} \! -\left(\frac{\bsc(\xx)}{\overline{\bsc}(\xx)}\right)^{\!\!\!\frac{ \bsbeta_r-\bsalpha_r}{2}}\right] (\bsbeta_r - \bsalpha_r) \\
	&= \sum\limits_{r=1}^R \kappa_r(\xx) \left[ \left(\frac{\bsc(\xx)}{\overline{\bsc}(\xx)}\right)^{\! \bsalpha_r} \! - \left(\frac{\bsc(\xx)}{\overline{\bsc}(\xx)}\right)^{\!\bsbeta_r}\right] (\bsbeta_r - \bsalpha_r)  \\
	&=  \sum\limits_{r=1}^R  \left(k_{+r} \, \bsc(\xx)^{ \bsalpha_r} -k_{-r} \, \bsc(\xx)^{\bsbeta_r}\right) (\bsbeta_r - \bsalpha_r) \\
	&= \sum\limits_{r=1}^R \mathfrak{r}^r(\bbc)(\xx)(\bsbeta_r - \bsalpha_r) \, .
\end{align*}

\subsubsection*{Diffusion part}

Given the second line of \eqref{dissipationRDPDE}, we set 
$F^{\text{diff}}_{\bbc}(\bbmu)  = \int_{\mathbb{T}^d}f^{\text{diff}}_{\bbc}(\xx,\bsmu(\xx),\nnabla \bsmu(\xx))\, \di \xx$ with 
\[f^{\text{diff}}_{\bbc}(\xx,\bsy,\bsz) = \frac{1}{2}  \sum\limits_{s=1}^S c^s(\xx) \bigl(z^s\bigr)^\intercal D^s(\xx) z^s \, . \]
This time, we have that $\frac{\partial}{\partial\bsy} f^{\text{diff}}_{\bbc}=0$,
while 
\[ \frac{\partial}{\partial \bsz}f^{\text{diff}}_{\bbc}(\xx,\bsy,\bsz)  =   (c^s(\xx) \, D^s(\xx) \, z^s)_{s=1,...,S} \in \mathbb{R}^{d \times S} \, ,\]
where $z^s\in \mathbb{R}^d, D^s(\xx) \in \mathbb{R}^{d \times d}$. 
Replacing $\bsz$ by $\nnabla \bsmu(\xx) $ and inserting $-\frac{\delta \mathfrak{e}}{\delta \bbc}(\bbc)$ for $\bbmu$, we get
\begin{align*}
	\frac{\delta F^{\text{diff}}_{\bbc}}{\delta \bbmu}\left(-\frac{\delta \mathfrak{e}}{\delta \bbc}(\bbc)\right)(\xx) &\; = \; \Biggl( \Div \biggl(c^s(\xx) \, D^s(\xx) \, \nnabla\log\frac{c^s(\xx)}{\overline{c}^s(\xx)} \biggr) \Biggr)_{s=1,...,S} \\
	&\hspace{-0.2cm} \overset{\eqref{DDDB}}{=} \Div\bigl(\bsD(\xx) \, \nnabla \bsc(\xx)\bigr) \\
	& \; = \;  -\Div \mathfrak{d}^s(\bbc)(\xx) \, .
\end{align*}
In combination with the reaction part, we obtain the RDPDE \eqref{RDPDE}.

\end{document}